\documentclass[prapplied,twocolumn,showpacs,preprintnumbers,amsmath,amssymb]{revtex4-2}

\usepackage{graphicx}
\usepackage{float}
\usepackage{xcolor}
\usepackage{bm} 
\usepackage[utf8]{inputenc}
\usepackage[T1]{fontenc}
\usepackage[%
  colorlinks=true,
  urlcolor=black,
  linkcolor=blue,
  citecolor=blue]{hyperref}
\usepackage[all]{hypcap}
\usepackage{sidecap}
\sidecaptionvpos{figure}{t}
\usepackage[capitalise]{cleveref}
\graphicspath{{pictures/}}
\usepackage{siunitx}
\usepackage{nicefrac}

\begin{document}

\title{Computational design of tunnel diodes with negative differential resistance and ultrahigh peak-to-valley current ratio based on two-dimensional cold metals: The case of NbSi$_2$N$_4$/HfSi$_2$N$_4$/NbSi$_2$N$_4$ lateral heterojunction diode}

% old title
%\title{Computational design of lateral heterojunction tunnel diodes utilizing 
%two-dimensional MA$_2$Z$_4$ (M=Nb, Ta; A=Si, Ge; Z=N, P) cold metals: Negative 
%differential resistance with high peak-to-valley current ratio}

\author{P. Bodewei$^{1}$}
\email{paul.bodewei@student.uni-halle.de}

\author{E. \c{S}a\c{s}{\i}o\u{g}lu$^1$}
\author{N. F. Hinsche$^1$}
\author{I. Mertig$^{1}$}

\affiliation{$^{1}$Institute of Physics, Martin Luther University Halle-Wittenberg, 06120 Halle (Saale), Germany} 

\date{\today}

\begin{abstract}

Cold metals have recently gained attention as a promising platform for innovative devices, 
such as tunnel diodes with negative differential resistance (NDR) and field-effect transistors 
with subthreshold swings below the thermionic limit. Recently discovered two-dimensional 
(2D) MA$_2$Z$_4$  (M=Ti, Zr, Hf, Nb, Ta; A=Si, Ge; Z=N, P) compounds exhibit both cold 
metallic and semiconducting behavior. In this work, we present a computational study of lateral 
heterojunction tunnel diodes based on 2D NbSi$_2$N$_4$ and HfSi$_2$N$_4$ compounds. Employing 
density functional theory combined with a nonequilibrium Green function method, we investigate 
the current-voltage ($I$-$V$) characteristics of lateral tunnel diodes with varying barrier 
thicknesses in both zigzag and armchair orientations. We find that tunnel diodes in the zigzag
orientation exhibit significantly higher peak current densities, while those in the armchair 
orientation display larger peak-to-valley current ratios (PVCRs) compared to the 
zigzag orientation. Our findings suggest that MA$_2$Z$_4$ materials are promising candidates 
for realizing NDR tunnel diodes with ultra-high PVCR values, which could have potential applications 
in memory, logic circuits, and other electronic devices.

\end{abstract}

\maketitle

\section{Introduction}

Negative differential resistance (NDR) tunnel diodes offer unique functionalities 
and potential applications when integrated within conventional complementary 
metal-oxide-semiconductor (CMOS) transistors \cite{berger2011negative}. 
These include NDR-based multi-valued logic gates, static random-access memory 
(SRAM), magnetic random-access memory, and low-power oscillators 
\cite{jo2021recent,wang2017leveraging,van1999tunneling}. 
Exploiting the NDR effect allows for logic and memory architectures with reduced 
device count, higher speed, and lower power consumption. For example, a tunnel 
SRAM requires only one transistor and two NDR tunnel diodes, instead of six 
transistors in a conventional SRAM. This results in a smaller footprint and 
lower power consumption \cite{van1999tunneling,karda2009one}. NDR diodes 
can be combined with CMOS transistors to implement various logic gates, while 
the extension to multi-valued logic increases information density and decreases 
system complexity \cite{micheel1990differential,lin1994resonant,jin2004tri}.

The NDR effect has been demonstrated in a variety of devices and circuits, with 
a focus on two-terminal tunnel diodes like Esaki diodes and resonant tunneling 
diodes \cite{esaki1958new, esaki1970superlattice, ramesh2012high, bruce2020insights}.
Esaki diodes, which operate via quantum tunneling under forward bias, show NDR behavior. 
However, the application of these two-terminal NDR diodes faces challenges. They often 
have low peak-to-valley current ratios (PVCR), which are not suitable for memory 
applications like tunnel SRAMs \cite{ramesh2012high,fung2011esaki,duschl1999high,schmid2012silicon}. 
Additionally, III-V semiconductors, which offer high PVCR, are difficult to integrate 
within current CMOS technology \cite{chow1992investigation,tsai1994pn}. Research 
efforts have sought to enhance PVCR values over 100 through CMOS-compatible processes \cite{van1999tunneling}. 
While some NDR circuits show extremely high PVCR values, their complex topology with 
multiple transistors makes them unsuitable for memory applications 
\cite{chung2004three,chen2009negative,gan2007fabrication,duane2003bistable,fang2022giant}.

\begin{figure}[!hb]
\begin{center}
\includegraphics[width=0.475\textwidth]{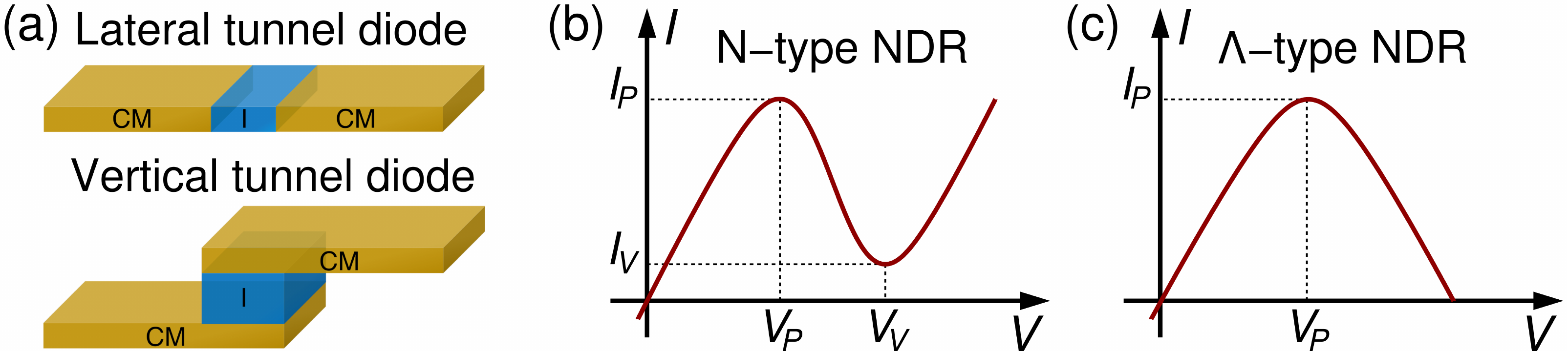}
\end{center}
\vspace*{-0.5cm} 
\caption{(a) Schematic representation of the lateral and vertical negative differential 
resistance (NDR) tunnel diodes based on cold metals. CM and I stand for cold metal and insulator, 
respectively. (b) The $I$-$V$ characteristics showing conventional N-type and  (c) $\Lambda$-type NDR.} 
\label{fig1}
\end{figure}

\begin{figure}
\begin{center}
\includegraphics[width=0.47\textwidth]{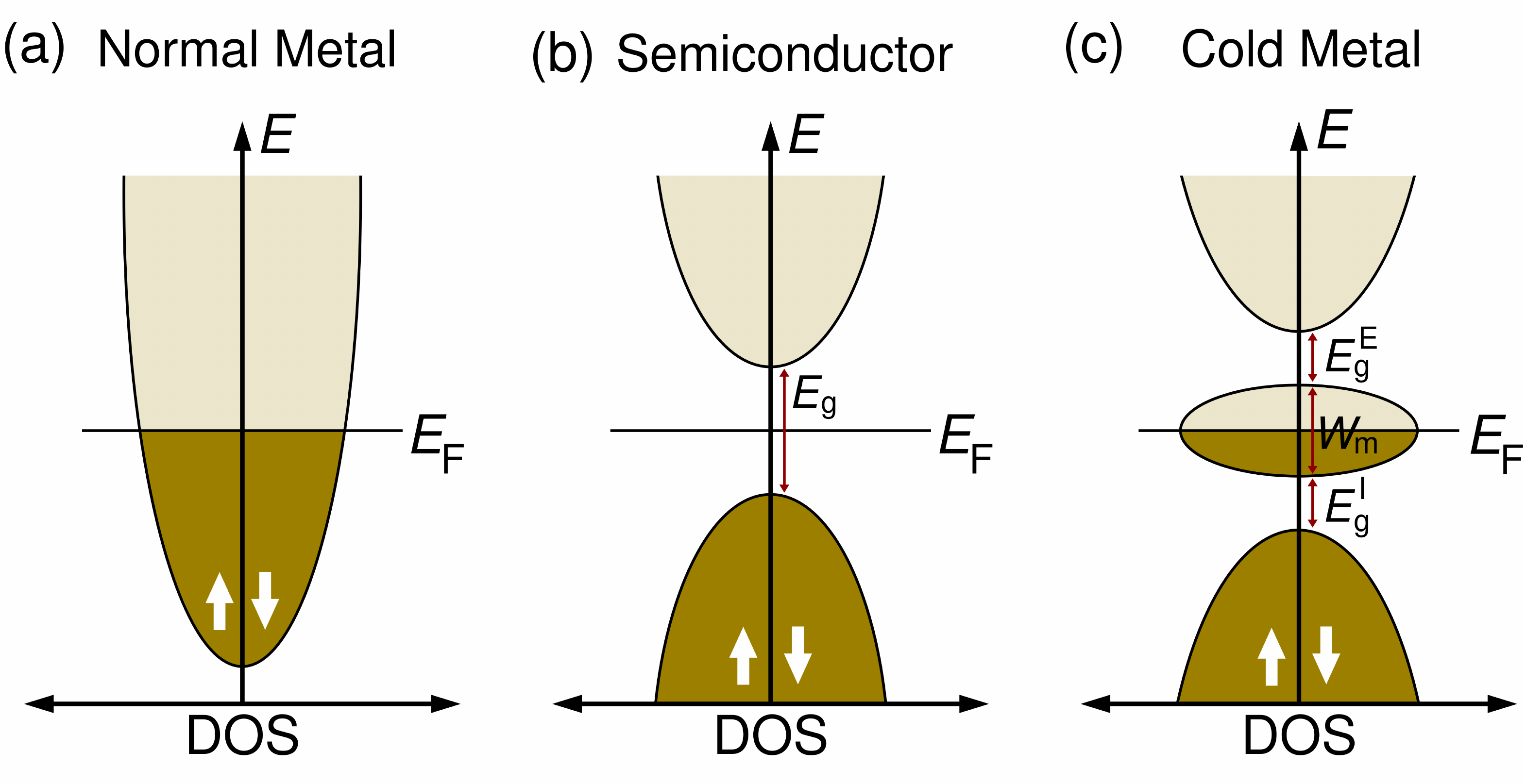}
\end{center}
 \vspace*{-0.65cm} 
\caption{Schematic illustration of the density of states (DOS) depicting: 
(a) a normal metal, (b) a semiconductor, and (c) a cold metal. The 
symbols $E_{\mathrm{g}}$, $E^{\mathrm{I}}_{\mathrm{g}}$, and $E^{\mathrm{E}}_{\mathrm{g}}$ 
correspond to the band gap of the semiconductor, as well as the internal and 
external band gaps of the cold metal, respectively. The width of the metallic 
band for the cold metal is designated by $W_{\mathrm{m}}$. The Fermi level is 
denoted by  $E_{F}$.} 
\label{fig2}
\end{figure}

Existing semiconductor-based  NDR tunnel diodes suffer from low peak-to-valley current ratio 
(PVCR) values, which is attributed to the band tails tunneling that originates from the strong 
doping and dopant fluctuations \cite{sant2017effect,bizindavyi2018band,schenk2020tunneling}.
To address this issue, we recently proposed a new semiconductor-free NDR tunnel diode concept with 
ultra-high PVCR \cite{sasioglu2023theoretical}. Our proposed diode consists of two cold metal 
electrodes separated by a thin insulating tunnel barrier (see Fig.\,\ref{fig1}). The NDR effect 
arises from the unique electronic band structure of the cold-metal electrodes (see Fig.\,\ref{fig2}). 
A cold metal is obtained when the metallic, i.e. partially filled, bands are separated from all 
energetically higher and lower-lying bands by sufficiently large energy gaps.
Specifically, the width of the isolated metallic bands around the Fermi level, as well as the energy 
gaps separating higher and lower-lying bands, determine the current-voltage ($I$-$V$) characteristics 
and PVCR of the tunnel diode. By choosing the appropriate cold 
metal electrode materials, either a conventional $N$-type or $\Lambda$-type NDR effect can be achieved 
as sketched in Fig.~\ref{fig1}(b,c).

The intention of this paper is to computationally design lateral heterojunction NDR tunnel 
diodes based on the recently discovered family of 2D materials known as MA$_2$Z$_4$ 
(where M=Ti, Zr, Hf, Nb, Ta; A=Si, Ge; Z=N, P) \cite{hong2020chemical}. These MA$_2$Z$_4$ 
compounds provide an exceptional platform for realizing NDR tunnel diodes due to their 
closely matched lattice parameters and composition, as well as their ability to exhibit 
both cold metallic and semiconducting properties within the same material 
class \cite{yin2023emerging}. This characteristic enables the coherent growth of consecutive 
components within the device which naturally favors a lateral device geometry \cite{latM}. 
Furthermore, we prefer a lateral heterojunction design towards the vertical counterpart, as a higher 
current density can be expected, especially pronounced in van der Waals 2D materials \cite{lateral_MoS2}.
An additional benefit of the MA$_2$Z$_4$ compounds is their superior 
strength and mechanical stability compared to most monolayer semiconductors like MoS$_2$ \cite{yin2023emerging}. 
Additionally, the electronic properties of MA$_2$Z$_4$ compounds are more resilient when 
it comes to stacking. For example, the cold metallic behavior observed in MX$_2$ (where M=Nb, 
Ta; X=S, Se, Te) is typically limited to monolayers, disappearing after only a few layers \cite{NFH_2018}. 
In contrast, MA$_2$Z$_4$ compounds maintain this behavior even in their bulk phase.

\section{Computational Method \label{sec:Method}}

The computational design of the presented lateral heterojunction tunnel diodes is based 
on density functional theory (DFT) within the QuantumATK T-2022.03 package \cite{smidstrup2020quantumatk}. 
We used a linear combination of atomic orbitals (LCAO) basis set (double zeta polarized) 
with norm-conserving FHI pseudopotentials \cite{troullier1991efficient}.  The exchange-correlation
functional was represented by a generalized gradient approximation (GGA) with Perdew-Burke-Ernzerhof
(PBE) flavor\cite{perdew1996generalized}. The self-consistent DFT calculations are performed using a 
$24 \times 24 \times 1$ \textbf{k}-point grid with a density mesh cut-off of $45 \, \mathrm{Hartree}$ 
and the total energy converged to at least $10^{-4} \, \mathrm{Hartree}$. To prevent interactions 
between the periodically repeated images, \SI{20}{\angstrom} of vacuum were added and Dirichlet 
and Neumann boundary conditions are employed. The transport calculations were performed using 
DFT combined with a nonequilibrium Green's function method (NEGF). We use a $24 \times 1 \times 184$ 
($14 \times 1 \times 318$)  $\mathbf{k}$-point grid in self-consistent DFT-NEGF calculations of the 
lateral tunnel diodes along the armchair (zigzag) orientation. The $I$-$V$ characteristics were 
calculated within a Landauer approach~\cite{Landauer-Buettiker}, where $ I(V) =\frac{2e}{h}\int 
\,T(E,V)\left[f_{L}(E,V)-f_{R}(E,V)\right] \mathrm{d}E $.  Here $V$ denotes the bias voltage, 
$T(E,V)$ are the transmission coefficients and $f_L(E,V)$ and  $f_R(E,V)$ are the Fermi-Dirac 
distributions of the left and right electrodes, respectively. 
The temperature was kept at $T=$~\SI{300}{K} throughout all calculations. 
The transmission coefficients $T(E,V)$ 
for the lateral tunnel diodes along the armchair (zigzag) orientation are calculated using a $255 
\times 1$ ( $147 \times 1$)  $\mathbf{k}$-point grid and convergence tests can be found in the 
supplemental material \cite{SI}.

\section{Results and Discussion \label{sec:ResandDis}}

In Fig.~\ref{fig1} we present schematically the structure of the lateral and vertical
NDR tunnel diode and the corresponding $I-V$ characteristics. The concept of the cold 
metal NDR tunnel diode was previously introduced by us in Ref.~\cite{sasioglu2023theoretical}, 
and thus only a brief overview of the device will be given subsequently. Our NDR tunnel 
diode consists of two cold metal electrodes, which are separated by an insulating tunnel 
barrier. The schematic density of states (DOS) of a cold metal is presented in 
Fig.~\ref{fig2} and compared with a DOS of a conventional metal and a semiconductor. As
seen in the schematic DOS cold metallic materials possess a unique band structure that
has an isolated metallic band $W_{\mathrm{m}}$ around the Fermi level as well as the 
energy gaps separating higher- and lower-lying states. The latter is referred to as
the internal gap $E^{\mathrm{I}}_{\mathrm{g}}$ (below the Fermi level) and  external 
$E^{\mathrm{E}}_{\mathrm{g}}$ (above the Fermi level). These three electronic structure
parameters play a decisive role in determining the $I$-$V$ characteristics of the tunnel
diode and the type of the NDR effect (see Fig.~\ref{fig1} (b)). A conventional $N$-type 
NDR effect is expected if these three parameters $W_{\mathrm{m}}$,  $E^{\mathrm{I}}_{\mathrm{g}}$, 
and $E^{\mathrm{E}}_{\mathrm{g}}$ are close to each other. On the other hand, the tunnel diode 
will show $\Lambda$-type NDR effect if $W_{\mathrm{m}} \ll E^{\mathrm{I}}_{\mathrm{g}} 
\sim E^{\mathrm{E}}_{\mathrm{g}}$.

\begin{table}
\caption{\label{tab:StructureParam}
Lattice constants $a_0$, work function $\Phi$, band gap $E_{\mathrm{g}}$, internal 
gap $E^{\mathrm{I}}_{\mathrm{g}}$, external gap $E^{\mathrm{E}}_{\mathrm{g}}$ and 
metallic bandwidth $W_{\mathrm{m}}$ of cold metallic and semiconducting 
MA$_2$Z$_4$ (M = Nb, Ta, Ti, Zr, Hf; A = Si, Ge; Z = N, P) monolayers. Lattice 
parameters are taken from Ref.\,\cite{yin2023emerging}. The band gap of HfSi$_2$N$_4$ 
calculated within the GGA+$U$ method is given in parentheses.}
\begin{ruledtabular}
\begin{tabular}{lcclccc}
Compound & $a_0$ & $\Phi$ &   $E_{\mathrm{g}}$ &  $E^{\mathrm{I}}_{\mathrm{g}}$ &  $E^{\mathrm{E}}_{\mathrm{g}}$ &  $W_{\mathrm{m}}$ \\
    & ({\AA}) &  (eV) & (eV)  &  (eV) & (eV) & (eV) \\ 
                \hline
NbSi$_2$N$_4$ & 2.97 &  5.62 &  & 1.01 & 1.74 & 1.33  \\
NbGe$_2$N$_4$ & 3.08 &  5.69 &  & 0.92 & 1.21 & 1.16  \\
NbSi$_2$P$_4$ & 3.53 &  4.52 &  & 0.01 & 0.52 & 1.26  \\
TaSi$_2$N$_4$ & 2.97 &  5.47 &  & 1.16 & 1.41 & 1.56  \\
TaGe$_2$N$_4$ & 3.08 &  5.51 &  & 1.16 & 0.98 & 1.25  \\
TaSi$_2$P$_4$ & 3.53 &  4.45 &  & 0.00 & 0.27 & 1.51  \\
\hline
TiSi$_2$N$_4$ & 2.94 &  6.08 & 1.54 &  &   & \\
ZrSi$_2$N$_4$ & 3.04 &  5.94 & 1.46 &  &   & \\
HfSi$_2$N$_4$ & 3.03 &  5.92 & 1.61 (1.84) &  &   & \\      
\end{tabular}
\end{ruledtabular}
\end{table}

\begin{figure}
\centering
\includegraphics[width=0.49\textwidth]{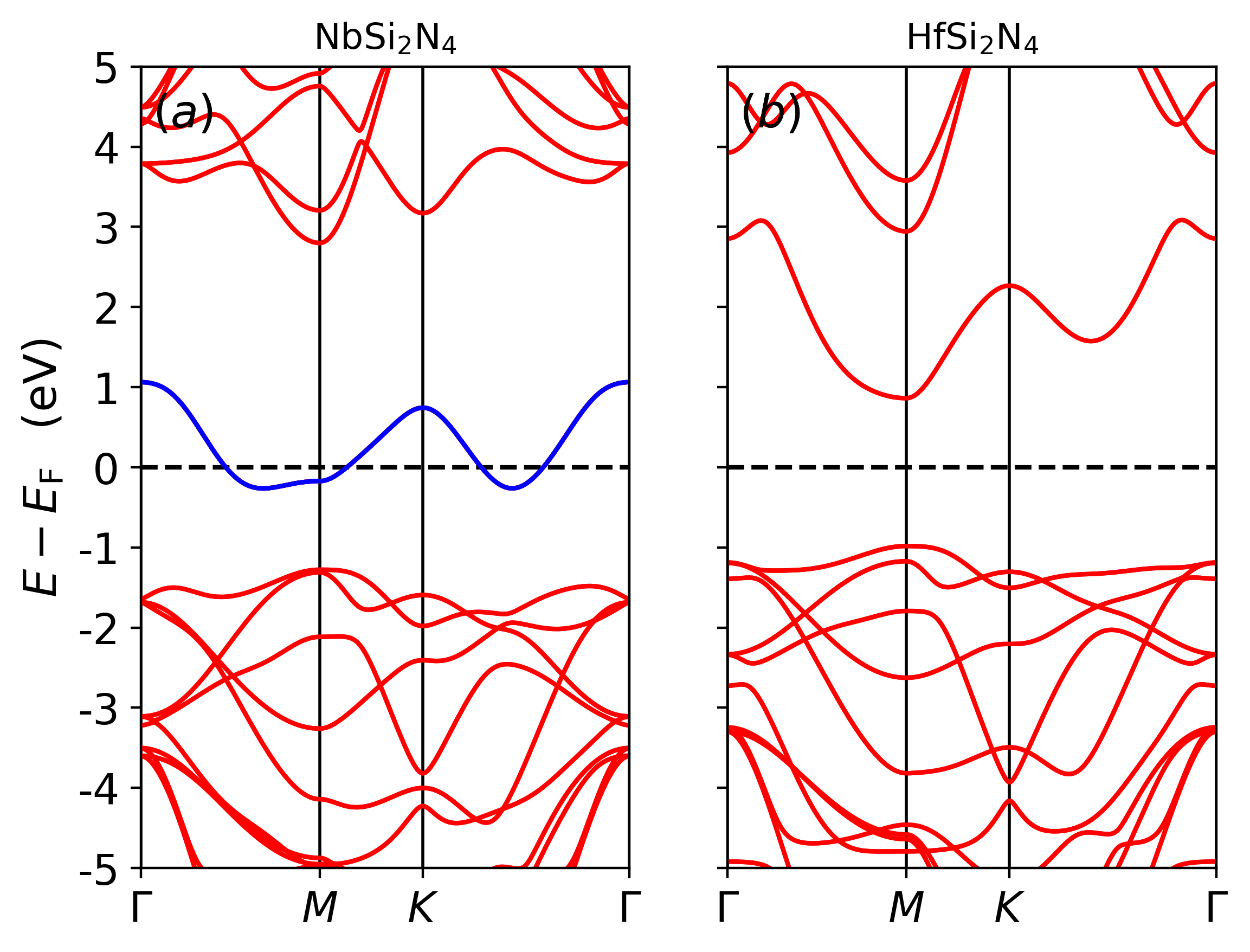}
\vspace*{-0.7cm}
\caption{Calculated PBE band structure of monolayer NbSi$_2$N$_4$ and PBE+$U$ band structure of 
HfSi$_2$N$_4$ along the high-symmetry lines in the 2D Brillouin zone. The dashed black lines 
denote the Fermi level, which is set to zero. The isolated metallic band of the NbSi$_2$N$_4$ 
is highlighted in blue. }
\label{fig:bandstructures}
\end{figure}

To design lateral NDR tunnel diodes, we first performed a material screening within the MoSi$_2$N$_4$ 
compound family, specifically focusing on the cold metal and semiconducting compounds, which are 
listed in Table~\ref{tab:StructureParam}. Our selection criteria for materials included cold metal 
electrode materials with large internal ($E_{\mathrm{g}}^{\mathrm{I}}$) and external 
($E_{\mathrm{g}}^{\mathrm{E}}$) band gaps relative to their metallic band widths ($W_{\mathrm{m}}$). For tunnel
barriers, we chose materials with large band gaps  ($E_{\mathrm{g}}$) and work functions that 
matched those of the cold metallic compounds. In Table~\ref{tab:StructureParam}, we provide 
the lattice parameters, work functions, band gaps ($E_{\mathrm{g}}^{\mathrm{I}}$, $E_{\mathrm{g}}^{\mathrm{E}}$, 
$E_{\mathrm{g}}$), and band widths for all the selected compounds. As presented, all the cold metal compounds 
under consideration, with few exceptions, have either internal or external band gaps smaller than their 
corresponding band widths, indicating that tunnel diodes based on these materials would exhibit the conventional 
$N$-type NDR effect. Among the screened materials, we selected NbSi$_2$N$_4$ as the cold metal electrode 
and HfSi$_2$N$_4$ as the tunnel barrier for our lateral tunnel diode simulations. The calculated electronic 
band structures of these materials are presented in Fig.~\ref{fig:bandstructures}. As all members 
of the 2D MoSi$_2$N$_4$ compound family belong to the trigonal $P\overline{6}m2$ ($\#187$) spacegroup, 
the bands are shown along the high-symmetry points of the related hexagonal Brillouin zone. 
The monolayer NbSi$_2$N$_4$ possesses an internal and external band gap of \SI{1.01}{eV} and \SI{1.74}{eV} 
respectively, while its metallic band width is \SI{1.33}{eV} within DFT-PBE. The tunnel barrier HfSi$_2$N$_4$ 
has a band gap of \SI{1.61}{eV} within DFT-PBE. Note that DFT-PBE underestimates the band gap of 
semiconductors and insulators, as well as cold metals (including cold metals such as the MX$_2$ 
(M=Nb, Ta; X=S, Se, Te) compounds \cite{heil2018quasiparticle,kim2017quasiparticle}). To improve the 
band gap of the tunnel barrier HfSi$_2$N$_4$, we employed a PBE+$U$ method with a $U$ value of \SI{8}{eV} for the 
$d$ orbitals of Hf in the transport calculations (see Table \ref{tab:StructureParam}).

\begin{figure}
\centering
\includegraphics[width=0.499\textwidth]{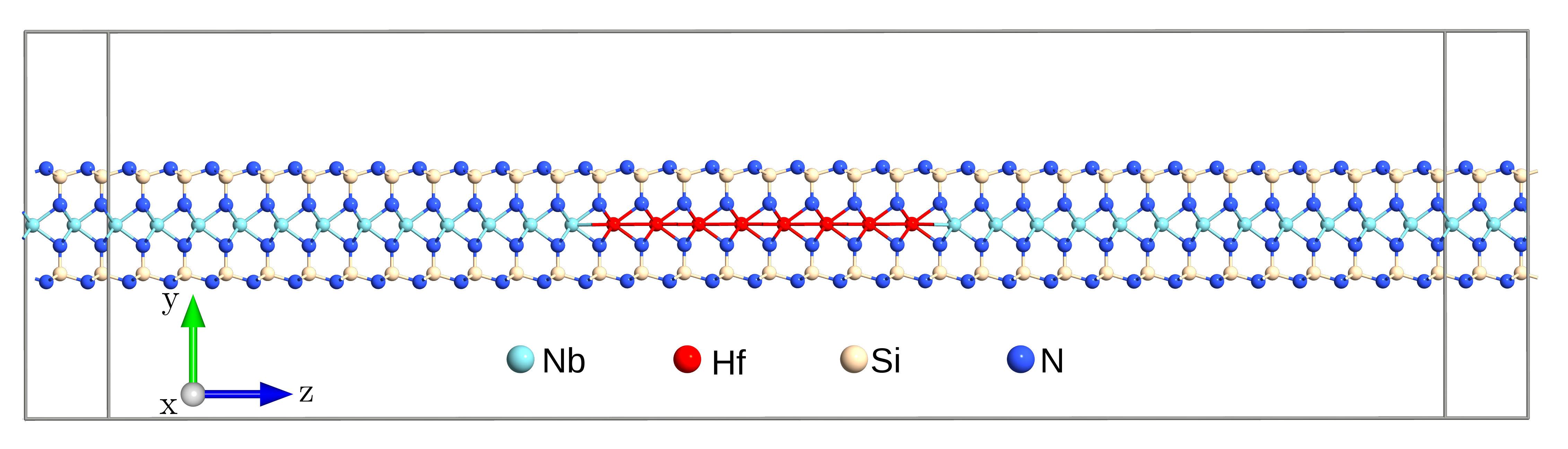}
\vspace*{-0.6cm}
\caption{Schematic illustration of the atomic structure of the NbSi$_2$N$_4$/HfSi$_2$N$_4$/NbSi$_2$N$_4$
lateral heterojunction tunnel diode device in armchair orientation. The tunnel barrier 
width is \SI{21}{\angstrom}. Different atomic components are represented by distinct colors.}
\label{fig3}
\end{figure}

In Fig.~\ref{fig3}, we show the atomic structure of the lateral tunnel diode, which is formed by attaching 
one monolayer of cold metal NbSi$_2$N$_4$ as the left and right electrodes and one monolayer of HfSi$_2$N$_4$ 
as the tunnel barrier. We consider both armchair and zigzag directions as the transport direction for the 
tunnel diodes, which we will refer to as armchair and zigzag tunnel diodes, respectively. The device is 
periodic in the $x$-direction, and we choose the $z$-direction as the transport direction. Since the 
lattice constants of the electrode and tunnel barrier materials are slightly different (less than $2\%$ 
mismatch, see Table\,\ref{tab:StructureParam}) for both directions, the tunnel barrier adopts the in-plane 
($x$-direction) lattice constant of the electrode material, and its lattice constant 
along the transport direction ($z$-direction) is relaxed. The tunnel barrier thicknesses are chosen to 
be about \SI{21}{\angstrom} and \SI{30}{\angstrom} for both armchair and zigzag tunnel
diodes. The total length of the scattering region ranges from  \SI{81}{\angstrom} to  \SI{90}{\angstrom}.

\begin{figure*}[t]
\centering
\includegraphics[width=0.95\textwidth]{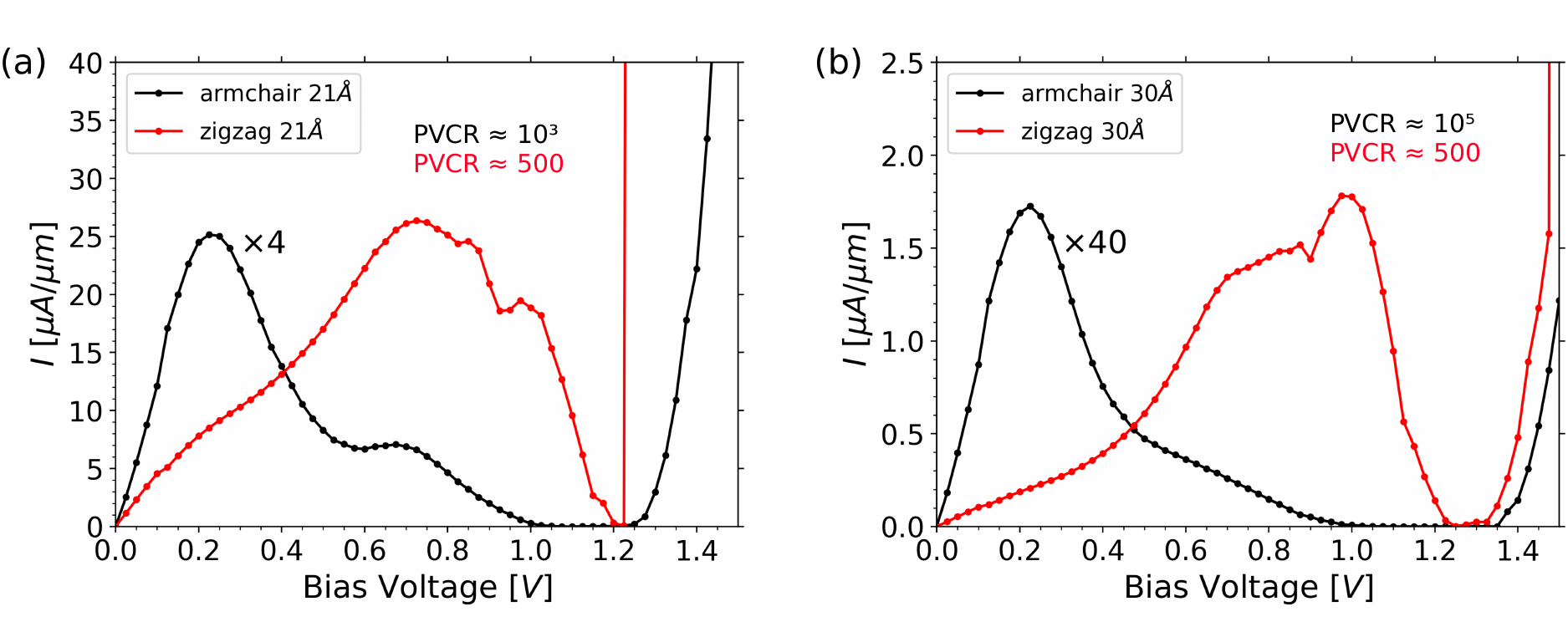}
\vspace*{-0.45cm}
\caption{(a) Calculated room temperature current-voltage characteristics for the  NbSi$_2$N$_4$/HfSi$_2$N$_4$/NbSi$_2$N$_4$ 
lateral heterojunction tunnel diode with \SI{21}{\angstrom} tunnel barrier in both armchair and zigzag 
orientations. (b) The same as (a) for a tunnel barrier length of \SI{30}{\angstrom}.}
\label{fig:21A_I-V}
\end{figure*}

In Fig.~\ref{fig:21A_I-V}, we present the calculated $I$-$V$ curves for tunnel diodes with tunnel barrier 
widths of \SI{21}{\angstrom} and \SI{30}{\angstrom}, for both armchair (black lines) and zigzag (red lines) 
directions. A maximum bias voltage of \SI{1.5}{V} was chosen. We report a conventional $N$-type NDR effect 
in both transport directions with a valley voltage $V_V$ of approximately \SI{1.25}{V}, which is primarily 
set by the metallic band width of the cold metal electrodes (see Table\,\ref{tab:StructureParam}). Besides 
this similarity, the $I$-$V$ curves of the armchair and zigzag directions exhibit distinct behavior. 
In the armchair case, the peak current $I_P$ (cf. Fig.~\ref{fig2} )is achieved at very low bias voltages 
of about \SI{0.2}{V}, while in the zigzag case it appears at higher bias voltages of $V>\SI{0.7}{V}$. 
These details can be directly attributed to the overlap of contributing electronic states of the left and 
right cold metal electrodes in different areas of the Brillouin zone. The mentioned overlap differs for 
armchair and zigzag-oriented tunnel diodes resulting in the presented $I$-$V$ curves in Fig.~\ref{fig:21A_I-V}. 
In a first approximation, the spin-independent transmission through an insulating barrier of width $d$ at zero 
bias is $T(k_z,E)=t_L(k_z,E)\exp\bigl[-2\kappa(k_z,E)d\bigr] t_R(k_z,E)$ \cite{Belashenko_04}. Here $t_L(k_z,E)$ 
and $t_R(k_z,E)$ are the interface transmission functions of the left and right metallic leads, respectively. 
Assuming perfect symmetry matching of the wave functions at the interfaces, the product $t_L(k_z,E) \cdot t_R(k_z,E)$ coincides with the joint density of states of left and right electrodes. The remaining factor involves the exponential decay length $1/\kappa(k_z,E)$ of the wave function within the insulating barrier. The latter is directly connected to the complex band structure $E( k_z + \imath \kappa)$ of the barrier, which will be discussed later in the letter. We introduce further insights on the specific structure of the $I$-$V$ curves maxima via the energy- and state-resolved transmission, as well as projected interface band structures in the supplementary material \cite{SI}.

From Fig.~\ref{fig:21A_I-V} we furthermore identify vanishing valley current in a rather large bias 
voltage interval of \SI{1.0}{V} to \SI{1.3}{V} for armchair tunnel diodes at both barrier 
thicknesses. We note, that this is in contrast to conventional Esaki diodes where a vanishing valley 
current is barely obtainable. For the zigzag tunnel diodes, the transition from the NDR to second 
positive differential resistance (PDR) takes place in a very narrow voltage window. Note that for the 
first PDR and NDR regions the current is formed by intra-band tunneling, i.e. from the single Nb d-band 
in the NbSi$_2$N$_4$ electrodes, while in the second PDR region, the current is due to inter-band tunneling.

\begin{figure*}[t]
\centering
\includegraphics[width=0.999\textwidth]{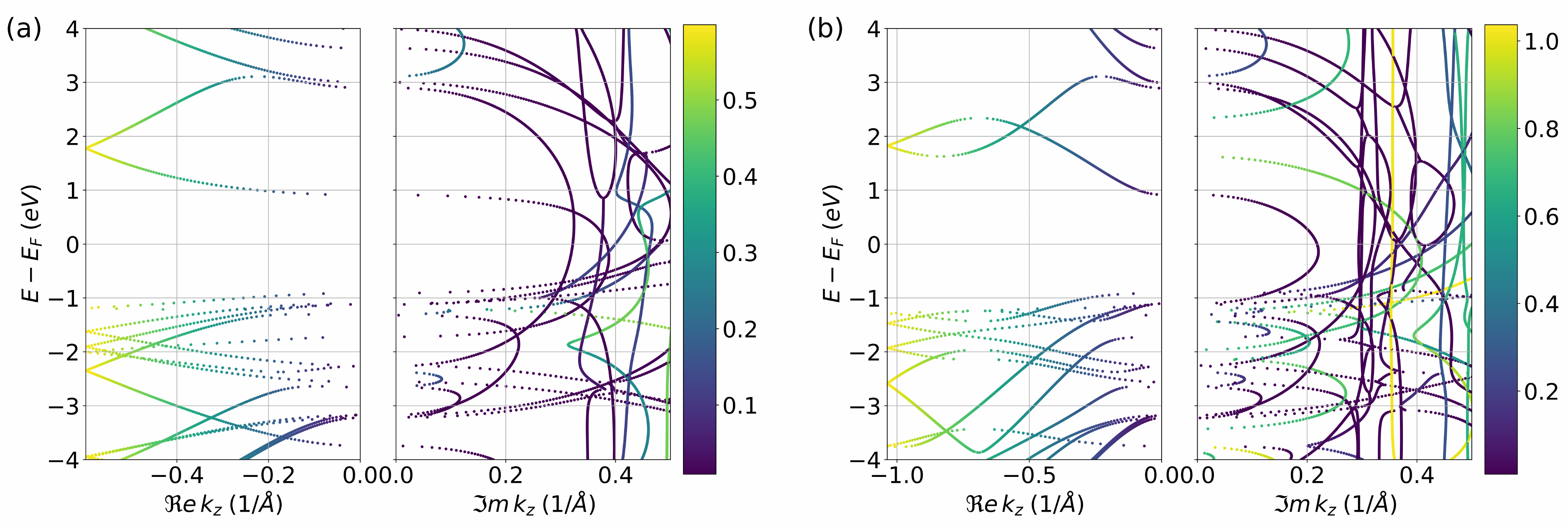}
\vspace*{-0.6cm}
\caption{Complex band structure of the barrier material HfSi$_2$N$_4$ in (a) armchair 
and (b) zigzag direction. The corresponding right panels show the imaginary parts 
$\kappa = \Im m k_z$ of the in general complex wavevector $k_z$. The colorbar reflects 
the value of the real part of $k_z$, i.e. purple bands indicate imaginary bands of the 
first kind with $\Re e k_z=0$; yellow bands correspond to imaginary bands of the second 
kind with $\Re e k_z=max$ and all other colors refer to complex bands with $\Re e k_z \neq 
0$ and $\Im m k_z \neq 0$.}
\label{fig:CBS}
\end{figure*}

The most striking difference is the absolute value of the peak current density in both device 
geometries. The maximum current density of the armchair tunnel diode located close to \SI{0.25}{V} is 
noticeably smaller than the peak current density of the zigzag tunnel diode located at $V$>\SI{0.7}{V}. 
In particular we find a current density ratio of $\nicefrac{I_{zz}}{I_{arm}}\approx 4$ for the shorter 
\SI{21}{\angstrom} device and $\nicefrac{I_{zz}}{I_{arm}}\approx 40$ for the longer \SI{30}{\angstrom} 
device. This difference can be readily explained with the help of the complex band structure 
$E(k_z+\imath\kappa)$, whereas $\kappa = \Im m k_z$. The latter is displayed for the tunnel barrier 
material HfSi$_2$N$_4$ for both devices orientations in Fig.~\ref{fig:CBS}. We recall that the 
transmission $T(k_z,E)=t_L(k_z,E)\exp\bigl[-2\kappa(k_z,E)d\bigr] t_R(k_z,E)$ through an insulating 
barrier is to a good approximation only dependent on the available state overlap of the left and right 
electrodes, their interface symmetry matching and the exponential decay length $1/\kappa(k_z,E)$ within 
the insulating barrier. It is worth mentioning, that in the case of the orthorhombic transport geometry 
of the explicit diode devices, the available states of the electrodes as well as barrier material within 
the range of the applied bias voltages consist only of $A'$, $A''$ symmetry characters of the $C_{1h}$ 
group, i.e. states symmetric or anti-symmetric with respect to reflection through the Hf/Nb mirror 
plane \cite{grouptheory_86,Dresselhaus_08}. As a consequence, there is almost no state-filtering at the interface 
and the electronic transport can be well understood by an analysis of the complex band structure. 
Obviously states with a small imaginary part $\kappa$ will have the weakest decay within the HfSi$_2$N$_4$ 
barrier and will dominate the contribution to the transmission, i.e. the current density. 

From Fig.~\ref{fig:CBS} is seen that the fundamental band gap for real-valued $k_z$ is bridged by 
loops with imaginary-valued $\kappa = \Im m k_z$. Comparing the complex band structures of 
the zigzag and armchair-oriented barrier material we immediately conclude that a single 
loop will dictate the transport in the zigzag geometry, while a combination of at least two 
crossing loops will describe the tunneling in the case of the armchair geometry. In all cases, 
imaginary loops have to connect real-valued $k_z$ of the same symmetry character \cite{NFH_TER}. 
The fact that an imaginary loop in the zigzag direction is considerably smaller than in the 
armchair direction directly reflects the reduced size of the current densities shown 
in Fig.~\ref{fig:21A_I-V}. For simplicity let $E=E_F$ and we find minimal values of 
$\kappa_{zz}\approx \SI{0.22}{1/\angstrom}$ and $\kappa_{arm}\approx \SI{0.32}{1/\angstrom}$ 
for the mid-gap states. Approximating the current density by
$I \approx \exp\bigl[-\kappa(k_z=0,E_F)d\bigr]$ we will obtain ratios of 
$\nicefrac{I_{zz}}{I_{arm}}\approx 8$ for the shorter \SI{21}{\angstrom} device and 
$\nicefrac{I_{zz}}{I_{arm}}\approx 20$ for the longer \SI{30}{\angstrom} device which 
is in excellent qualitative agreement with the results of our full NEGF calculations presented
in Fig.~\ref{fig:21A_I-V}. Due to the evanescent nature of the states the ratio
$\nicefrac{I_{zz}}{I_{arm}}$ will naturally increase for larger barrier widths.

While the peak current density in zigzag tunnel diodes is notably higher, it is essential to 
note that these exhibit smaller PVCR values. The PVCR values obtained for both tunnel barrier 
thicknesses are approximately 500. In contrast, for armchair tunnel diodes, the PVCR values 
are significantly higher, reaching $10^3$ and $10^5$ for tunnel barriers of \SI{21}{\angstrom} 
and \SI{30}{\angstrom}, respectively. These findings highlight the trade-off between peak 
current density and PVCR, and the unique characteristics of armchair and zigzag tunnel diodes 
for diverse applications. We note that these PVCR values are order of magnitudes higher than 
found in conventional Ge-based NDR diodes (PVCR~$\approx 2 \dots 10$), CaF$_2$-based NDR diodes 
(PVCR~$\approx 100$) \cite{berger2011negative} or even in MoS$_2$ 
homojunctions with a PVCR~$\approx 200 \dots 400$ \cite{lateral_MoS2}.

We want to emphasize that the choice of pseudopotentials and basis sets can be crucial in 
the evaluation of the complex band structure and thus of the related $I$-$V$ characteristics 
within the LCAO-NEGF formalism. Pay attention to the vertical complex bands appearing with an almost 
constant decay length $1/\kappa_z$ over a wide energy range - e.g. the imaginary band of 
second kind at $\kappa = \Im m k_z \approx \SI{0.35}{1/\angstrom}$ in the case of zigzag 
orientation shown in Fig.~\ref{fig:CBS}(b). These spurious states referred to as \textit{ghost} 
or \textit{phantom} modes, are associated with virtual molecular orbitals of the 
barrier area and might lead to an inaccurate description of the tunneling \cite{Herrmann_ghost}. 
Interestingly, the ghost modes are an apparently unavoidable consequence
of large numerical basis sets, so an allegedly increase in computational accuracy might 
increase the number of ghost states \cite{Herrmann_ghost}. A separation of ghost modes 
and \textit{true} CBS modes is challenging. In our case, the ghost states do have large $\kappa$, 
are heavily damped, and thus only contribute negligibly to the transport. Generally, they may 
exhibit very slow decay and can dominate the transport leading to fundamental qualitative and 
quantitative errors. We suggest using the complex band structure and its validity concerning 
the ghost states as a first test before the computationally demanding Landauer approach is 
applied.

The  $I$-$V$ characteristics and high PVCR values reported for the tunnel diodes are based on 
the assumption of coherent tunneling transport. This means that mechanisms like electron-electron 
scattering or electron-phonon scattering are not included in the calculations.
Electron-phonon scattering is a well-known dissipation mechanism, particularly efficient in 
2D semiconductors compared to 3D\cite{cheng2020two}. It can influence current flow in devices 
based on 2D materials \cite{afzalian2021ab}. While software packages like QuantumATK can include 
electron-phonon scattering, the computational cost is significant and often limited to very 
simple models\cite{smidstrup2019an}. Therefore, for our presented tunnel diodes, such detailed simulations 
are currently not practical. Consequently, the peak current in Fig.~\ref{fig:21A_I-V} and the PVCR 
obtained here represent upper limits within the coherent tunneling framework. Introducing inelastic 
scattering mechanisms, like electron-phonon interactions, is expected to modify the  $I$-$V$ characteristics. 
This would likely decrease the peak current and increase the valley current, resulting in a lower 
PVCR value. It is important to note that other factors, such as defects, interface atomic mixing, 
and interactions with surrounding materials, can also influence the  $I$-$V$ characteristics. However, 
the overall trends observed in the  $I$-$V$ curves are expected to remain largely unchanged.

Eventually, we discuss the anticipated $I$-$V$ characteristics of tunnel diodes based on other 
cold metals listed in Table~\ref{tab:StructureParam}. As mentioned, the $N$-type NDR effect 
is expected to be common in tunnel diodes using these materials. For example, TaSi$_2$N$_4$, 
which is isoelectronic to NbSi$_2$N$_4$, is expected to exhibit qualitatively similar $I$-$V$ 
curves. However, quantitative differences, such as peak current density, valley voltage, 
and valley current, are likely due to TaSi$_2$N$_4$'s larger metallic band width ($W_{\mathrm{m}}$) 
and slightly smaller internal and external band gaps. Similarly, Ge-based compounds are 
expected to exhibit analogous $I$-$V$ characteristics. On the other hand, tunnel diodes 
based on P-containing compounds like NbSi$_2$P$_4$ may exhibit significantly smaller PVCR 
values, primarily due to the vanishing internal band gap in these compounds. This insight 
into the expected behavior of tunnel diodes based on various cold metals offers a glimpse 
into the diverse possibilities for tailoring their performance in electronic applications 
and underscores the role of material properties in shaping device behavior.

\section{Conclusions}

In conclusion, our computational investigation of prototype lateral heterojunction tunnel diodes 
unveils the remarkable potential of cold metallic MA$_2$Z$_4$ (M=Nb, Ta; A=Si, Ge; Z=N, P) 
compounds for harnessing the NDR phenomenon with very high PVCR values orders of magnitude 
higher than in conventional NDR tunnel diodes. By considering the exemplary cold metal 
NbSi$_2$N$_4$ as the prototype electrode material and HfSi$_2$N$_4$ as the tunnel barrier, 
our calculations consistently demonstrate the achievement of $N$-type NDR behavior in 
both armchair and zigzag-oriented tunnel diodes. Moreover, our findings reveal intriguing 
differences between these orientations. Zigzag-oriented tunnel diodes exhibit significantly 
higher peak current densities, indicating their potential for high-speed electronic applications, 
while armchair-oriented diodes achieve very high PVCR values. This duality in performance 
highlights the versatility of MA$_2$Z$_4$ materials and their promising role in enabling 
future electronic devices with enhanced functionality and efficiency. Our findings not only 
broaden our understanding of the NDR effect in cold metal based heterojunction tunnel 
diodes, but also pave the way for exciting new directions in materials engineering for 
next-generation electronics.

\begin{acknowledgments}
This work was supported by SFB CRC/TRR 227 of Deutsche Forschungsgemeinschaft (DFG) and by the 
European Union (EFRE) via Grant No: ZS/2016/06/79307. 
\end{acknowledgments}

% \bibliography{PR_Applied_main_resub.bib}


\pagebreak
\onecolumngrid

\section*{Supplemental Material: 
Computational design of tunnel diodes with negative differential resistance and ultrahigh peak-to-valley current ratio based on two-dimensional cold metals: The case of NbSi$_2$N$_4$/HfSi$_2$N$_4$/NbSi$_2$N$_4$ lateral heterojunction diode}

P. Bodewei$^{1}$, E. \c{S}a\c{s}{\i}o\u{g}lu$^1$, N. F. Hinsche$^1$, I. Mertig$^{1}$
$^{1}$Institute of Physics, Martin Luther University Halle-Wittenberg, 06120 Halle (Saale), 
Germany

\begin{figure}[H]
    \centering
    \includegraphics[width=0.75\textwidth]{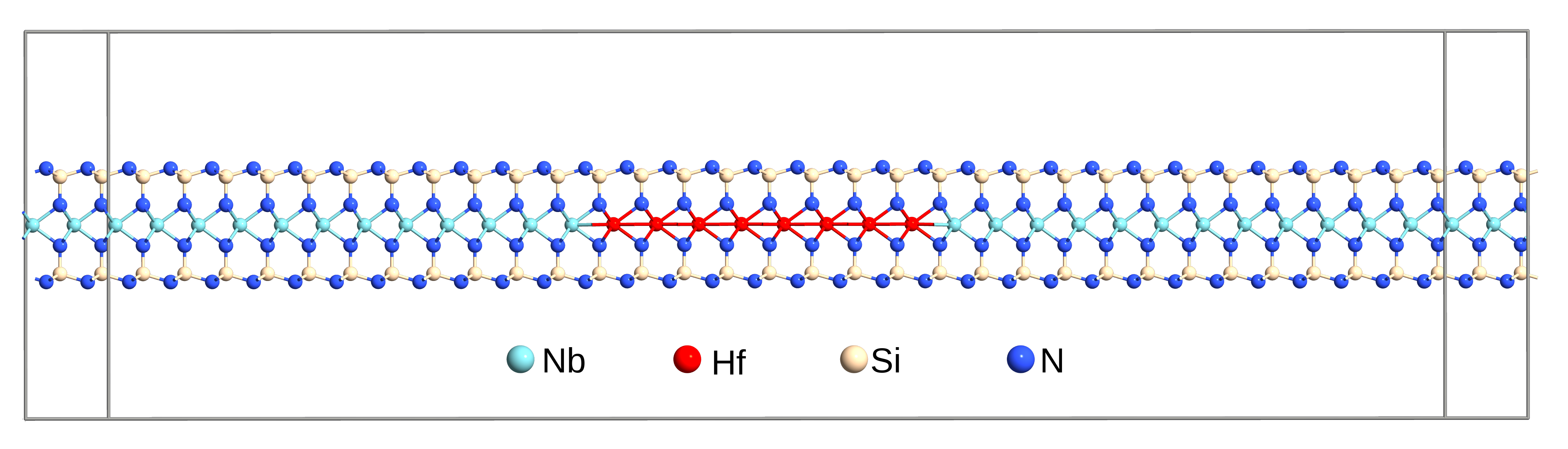}
    \includegraphics[scale=0.045]{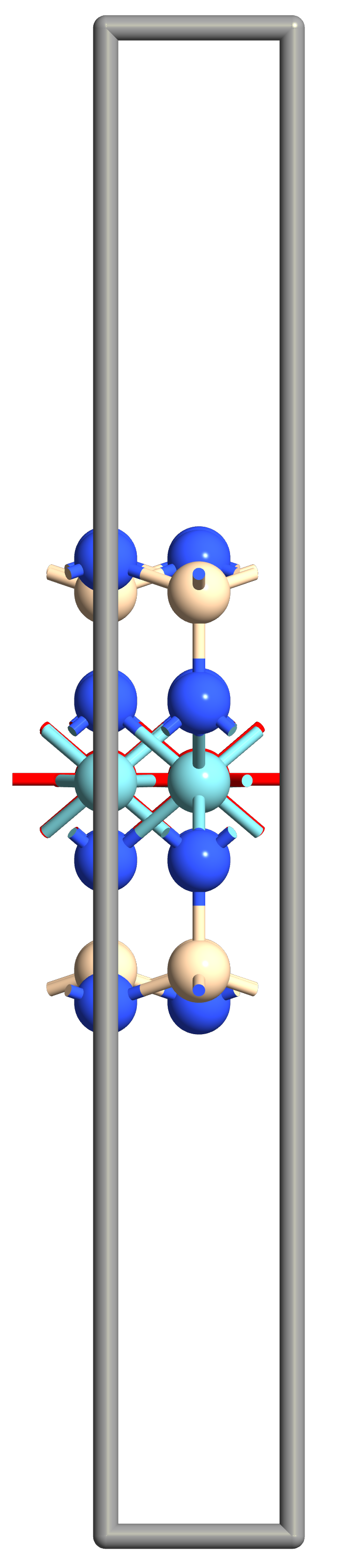}
    \caption{Schematic illustration of the atomic structure of the NbSi$_2$N$_4$/HfSi$_2$N$_4$/NbSi$_2$N$_4$
lateral heterojunction tunnel diode device in armchair orientation. The tunnel barrier 
width is \SI{21}{\angstrom}. The side view of the device is shown on the right side. 
Different atomic components are represented by distinct colors.}
    \label{fig:armchair-dev}
\end{figure}

\begin{figure}[H]
    \centering
    \includegraphics[width=0.75\textwidth]{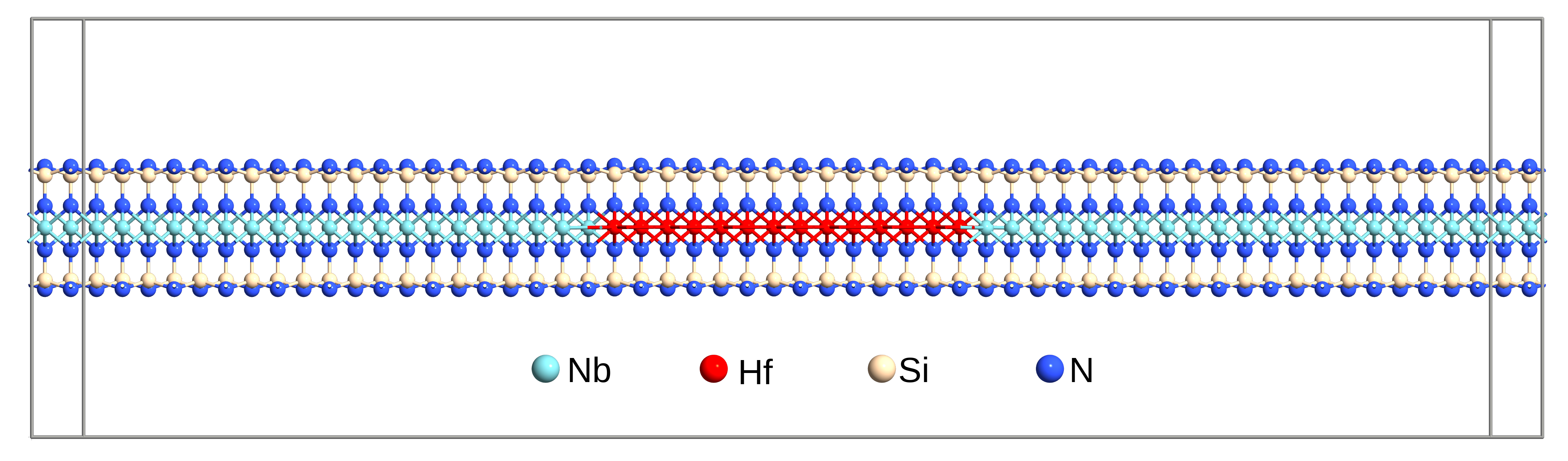}
    \includegraphics[scale=0.06]{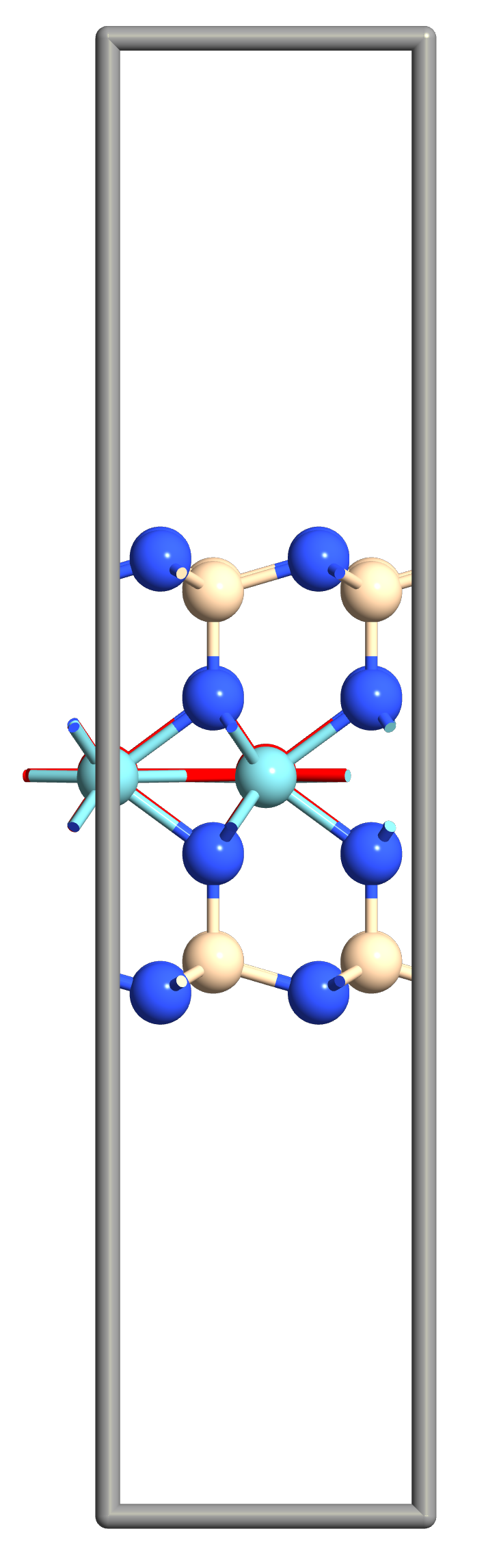}
    \caption{Schematic illustration of the atomic structure of the NbSi$_2$N$_4$/HfSi$_2$N$_4$/NbSi$_2$N$_4$
lateral heterojunction tunnel diode device in zigzag orientation. The tunnel barrier 
width is \SI{21}{\angstrom}. The side view of the device is shown on the right side. 
Different atomic components are represented by distinct colors.}
    \label{fig:zigzag-dev}
\end{figure}

\begin{figure}[H]
    \centering
    \includegraphics[width=0.75\textwidth]{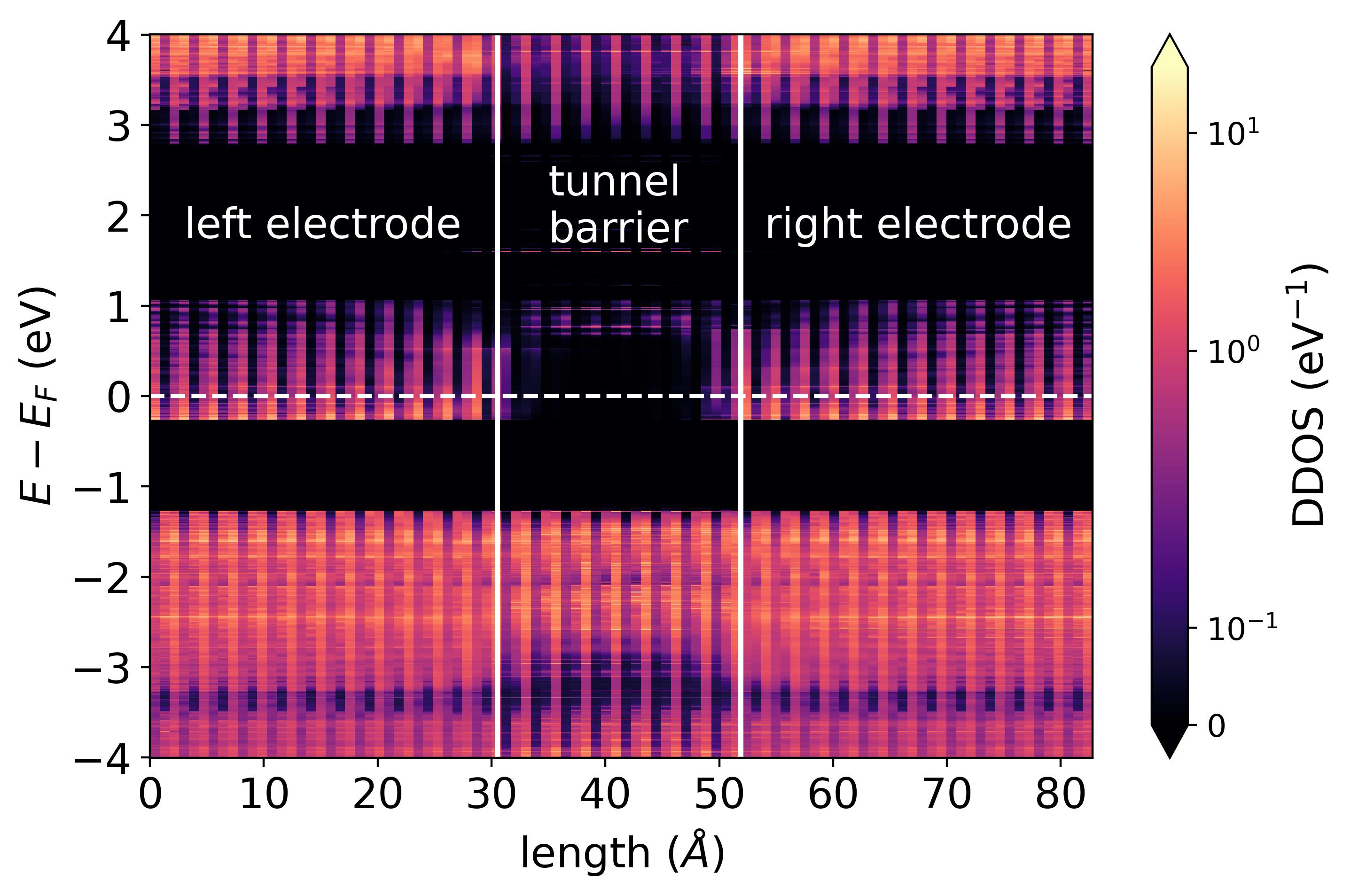}
    \vspace*{-0.5cm}
    \caption{Device density of states (DDOS) of the armchair device with the \SI{21}{\angstrom} barrier for a bias voltage of \SI{0}{V}.}
    \label{fig:PLDOS}
\end{figure}

\begin{figure}[H]
    \centering
    \includegraphics[width=0.6\textwidth]{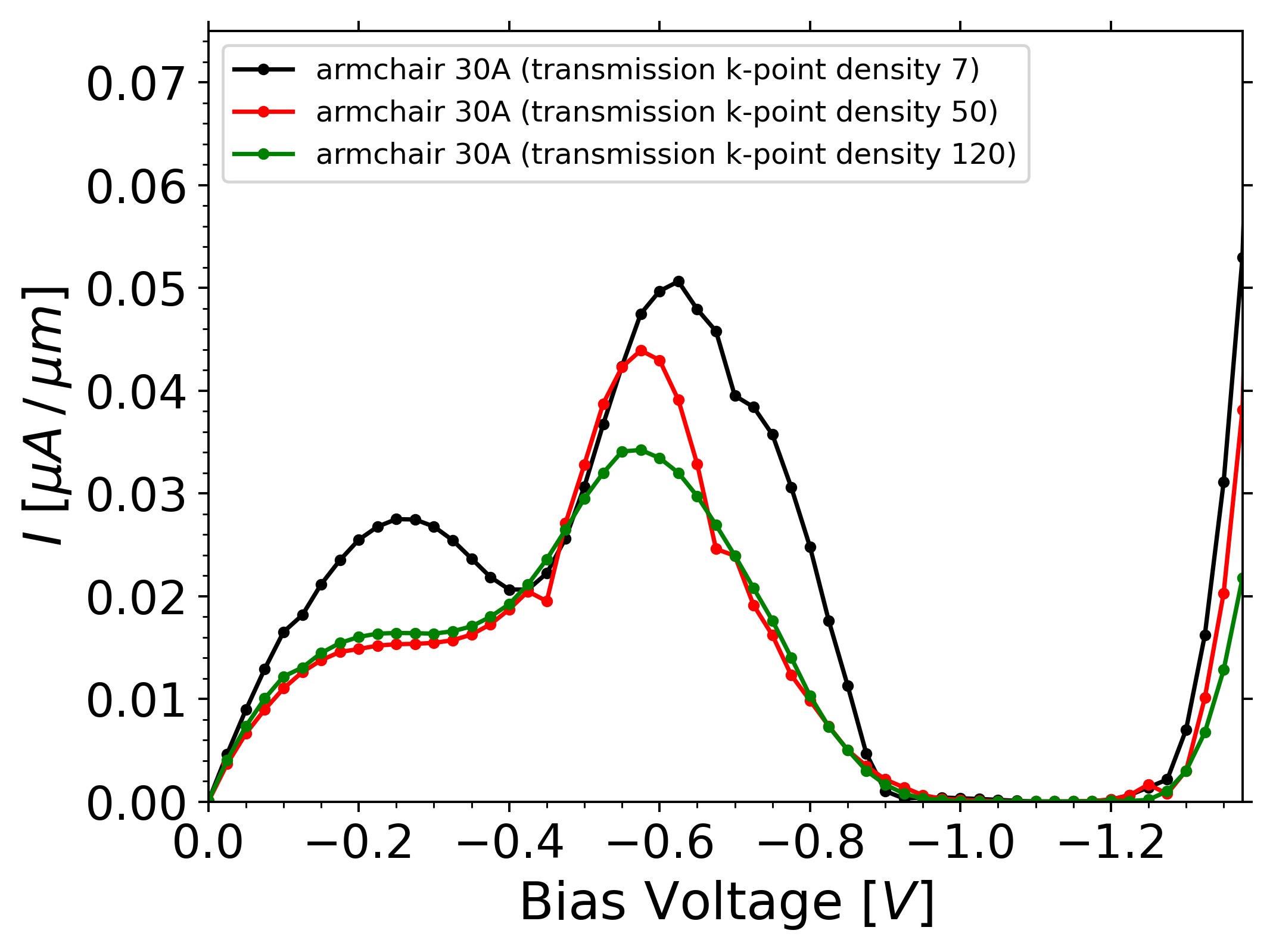}
    \caption{Convergence test for the $k$-point sampling for the calculation of the $I$-$V$-curve for the armchair device with \SI{30}{\angstrom} barrier thickness for negative bias applied with constant cut-offs, pseudo- and exchange-correlation-potentials}
    \label{fig:Convergence_kpoints}
\end{figure}

\begin{figure}[H]
    \centering
    \begin{minipage}{0.49\textwidth}
        \includegraphics[width=\textwidth]{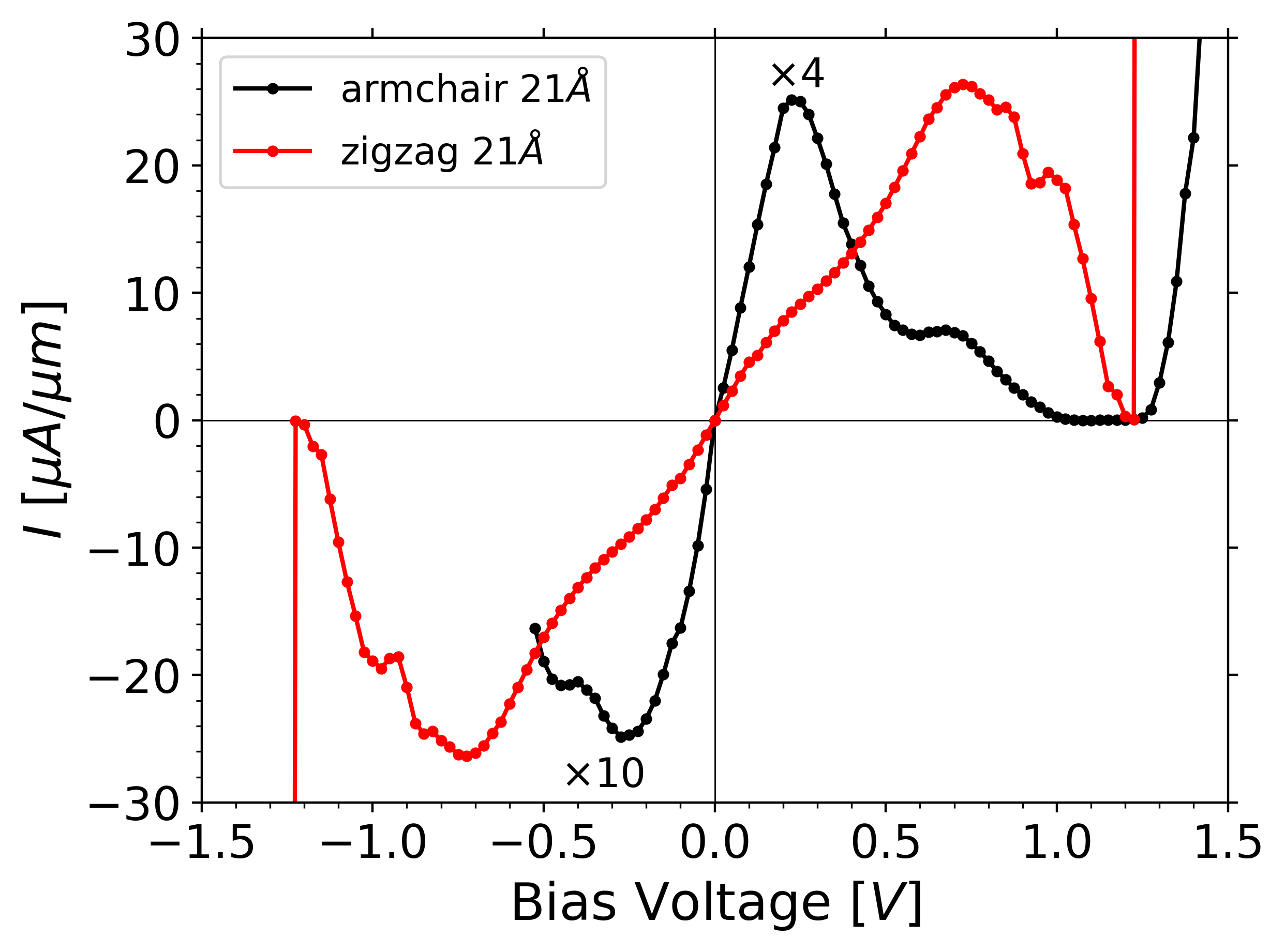}
        \caption*{(a) \SI{21}{\angstrom} barrier}
    \end{minipage}
    \hfill
    \begin{minipage}{0.49\textwidth}
        \includegraphics[width=\textwidth]{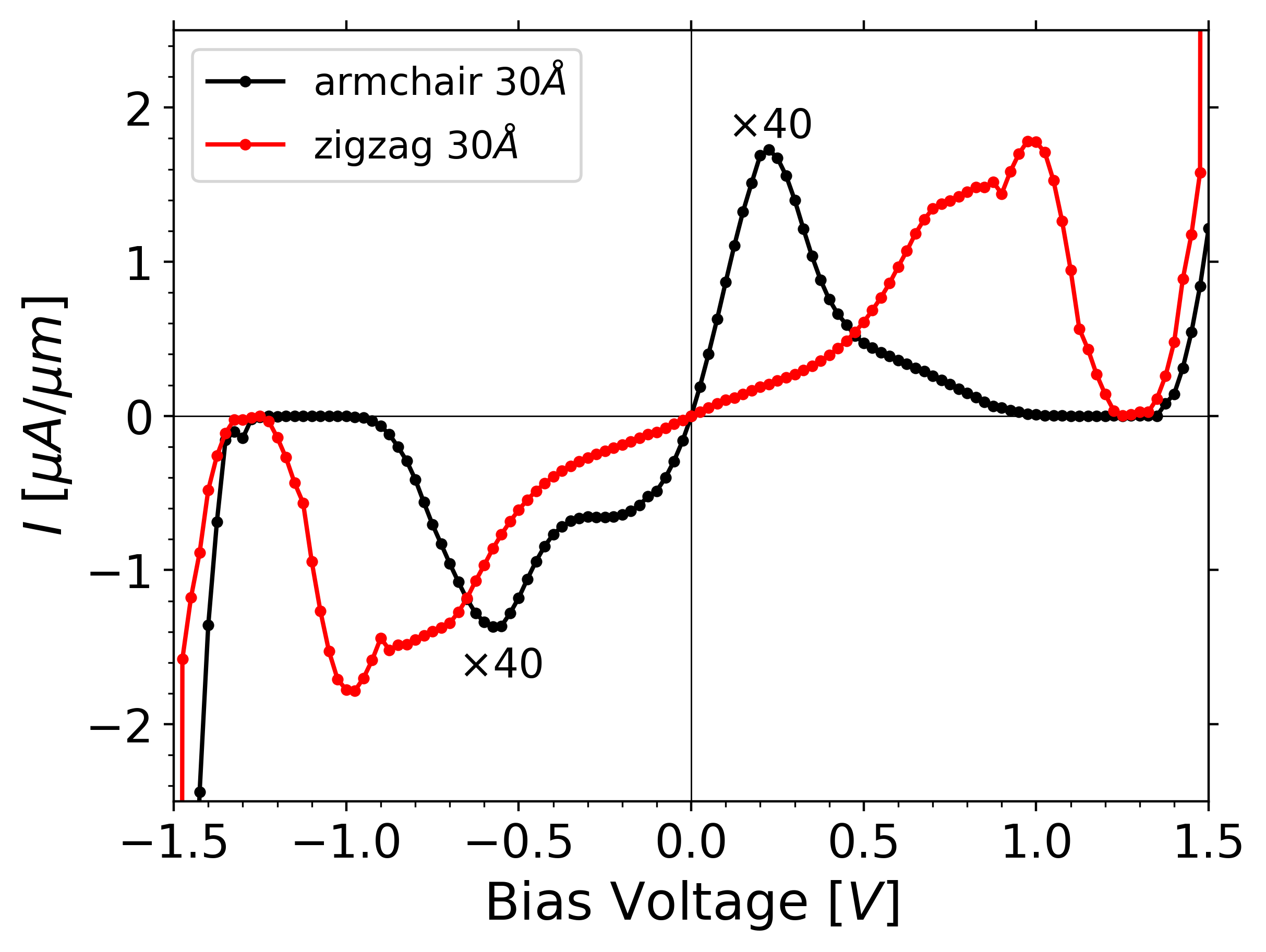}
        \caption*{(b) \SI{30}{\angstrom} barrier}
    \end{minipage}
    \caption{$I$-$V$-characteristics for positive and negative bias for the devices with \SI{21}{\angstrom} and \SI{30}{\angstrom} tunnel barrier \label{fig:suppl_fig5}
    }
\end{figure}

In Fig.~\ref{fig:suppl_fig5} the $I$-$V$-curves for positive and negative bias are shown. While the $I$-$V$-curve 
for the zigzag device with \SI{21}{\angstrom} and \SI{30}{\angstrom} barrier thickness are perfectly antisymmetric, 
we see that the $I$-$V$-curves for devices in armchair direction show a slight asymmetry. The reason for that is the interface potential, 
building up between NbSi$_2$N$_4$ and HfSi$_2$N$_4$. Geometrically, we can see that the termination of the interface 
between the right electrode and left electrode with the barrier is different in armchair devices. Thus, a different 
interface potential at both interfaces builds up, leading to different current densities for positive and negative 
bias voltages, respectively.

\begin{figure}[H]
    \centering
    \begin{minipage}{0.49\textwidth}
        \includegraphics[width=\textwidth]{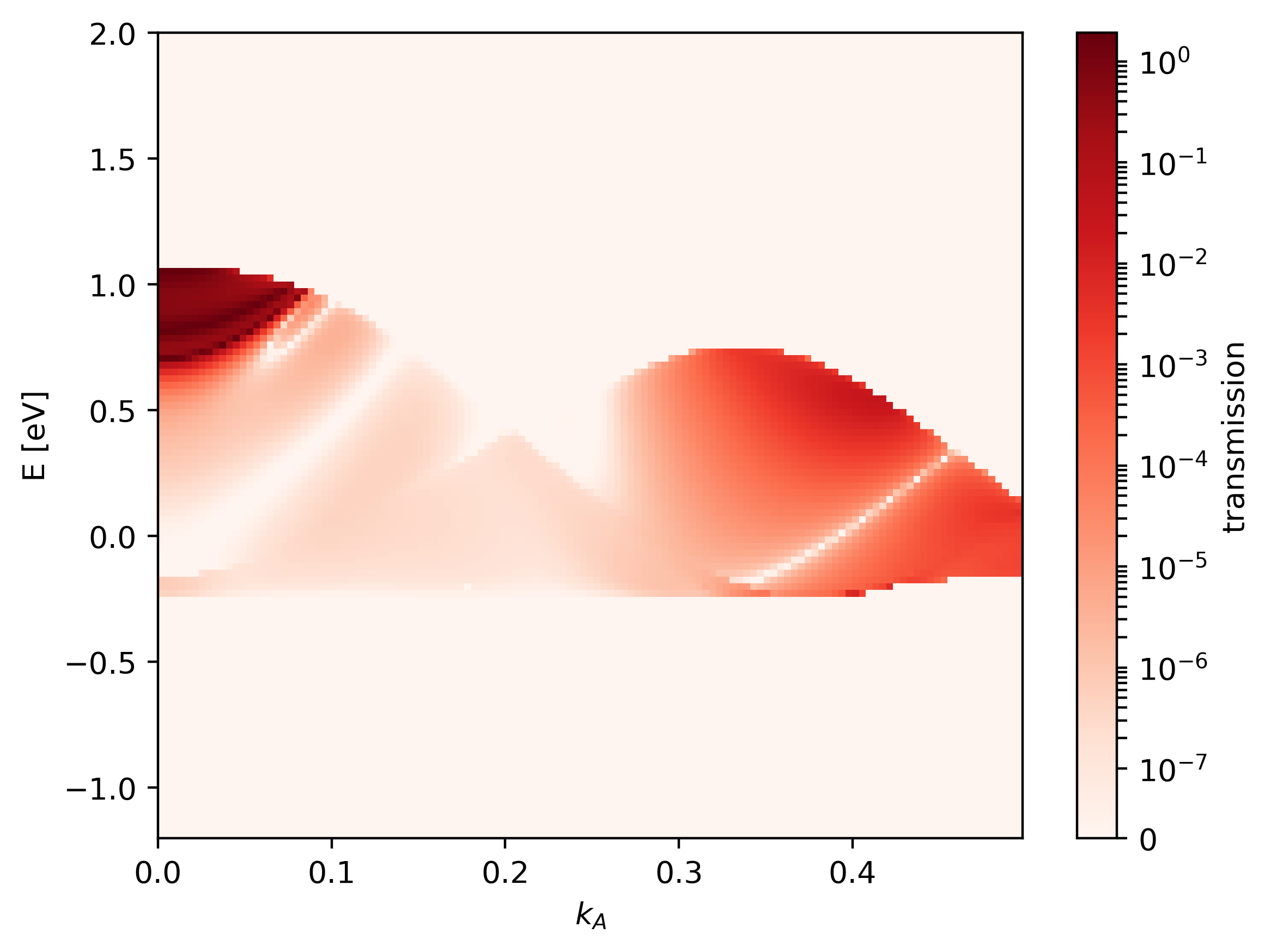}
        \caption*{(a) armchair device with \SI{21}{\angstrom} tunnel barrier}
    \end{minipage}
    \hfill
    \begin{minipage}{0.49\textwidth}
        \includegraphics[width=\textwidth]{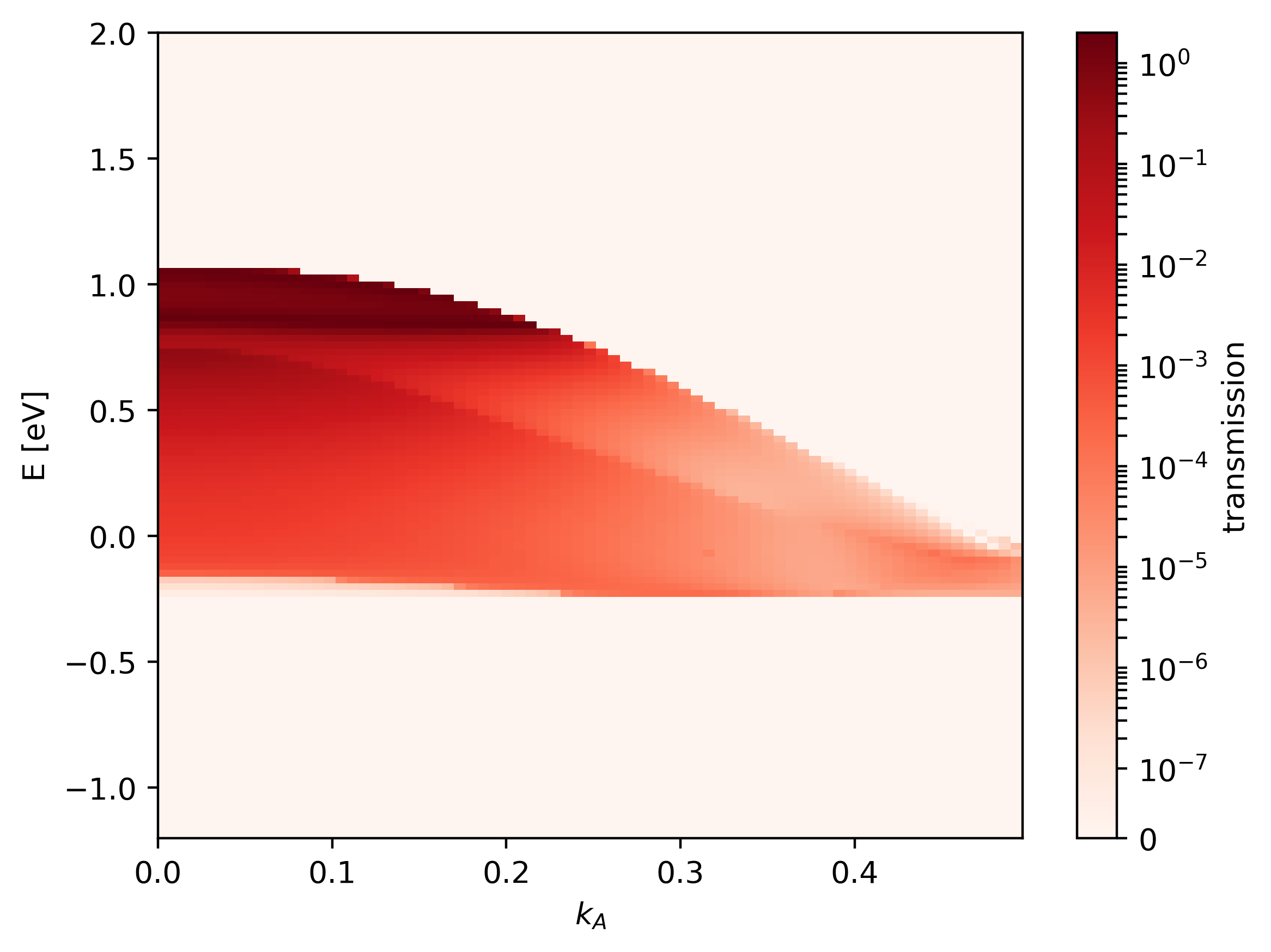}
        \caption*{(b) zigzag device with \SI{21}{\angstrom} tunnel barrier}
    \end{minipage}
    \caption{Transmission of the armchair and zigzag device for different energies and k-points in the direction perpendicular to the (a) armchair and (b) zigzag transport direction
    \label{fig:suppl_fig6}}
\end{figure}

In Fig.~\ref{fig:suppl_fig6} we see the transmission at different $k$-points for different energies. Firstly, we see that there is no certain $k$-point dominating the transmission, i.e. the transmission is distributed throughout the whole range of $k$-points for both the armchair and the zigzag device. Further, we see that for higher energies of about \SI{1}{eV} we see a strong contribution of the transmission at the $\Gamma$-point.

\begin{figure}[H]
    \centering
    \begin{minipage}{0.32\textwidth}
        \includegraphics[width=\textwidth]{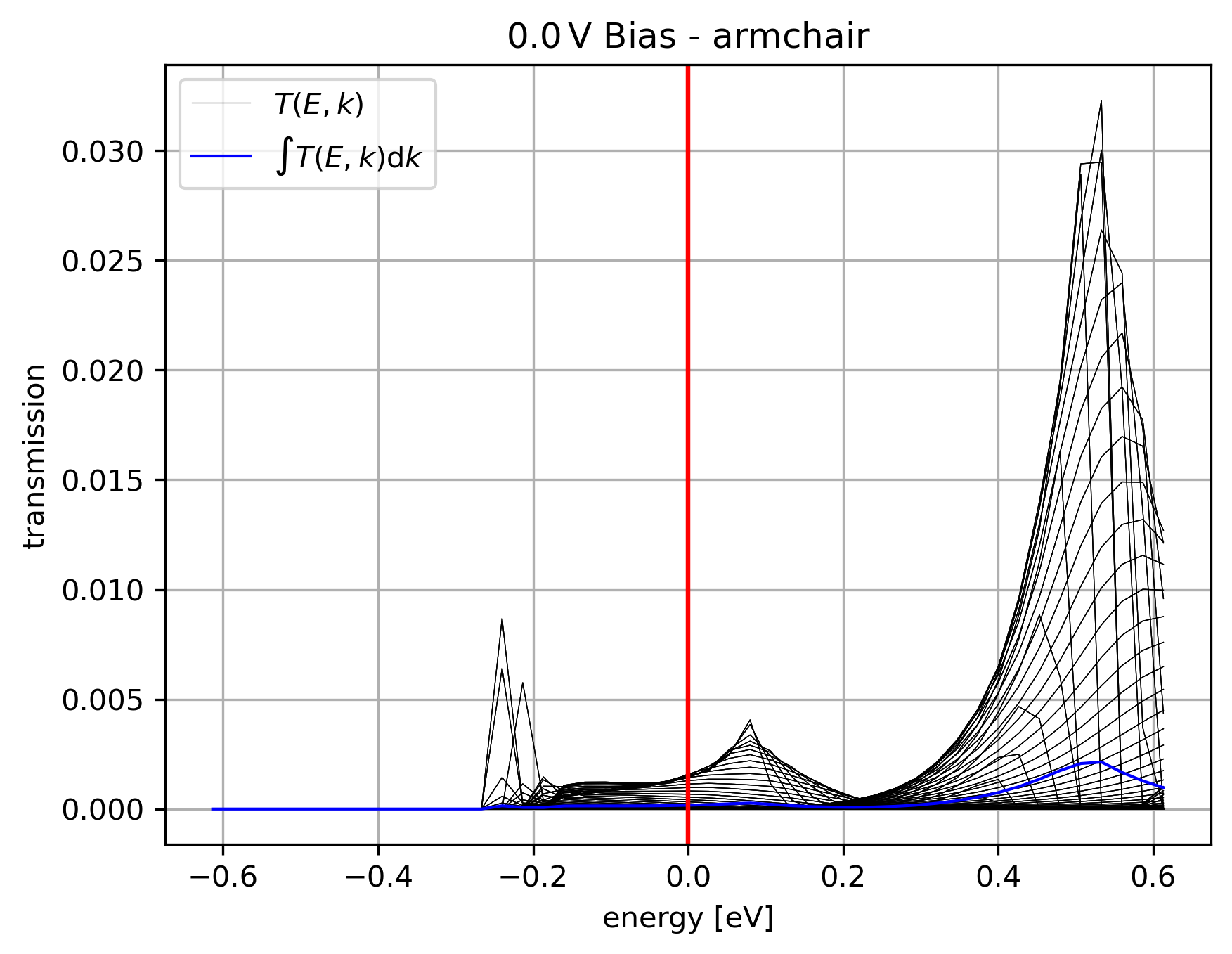}
        \caption*{\SI{0.0}{V} Bias}
    \end{minipage}
    \hfill
    \begin{minipage}{0.32\textwidth}
        \includegraphics[width=\textwidth]{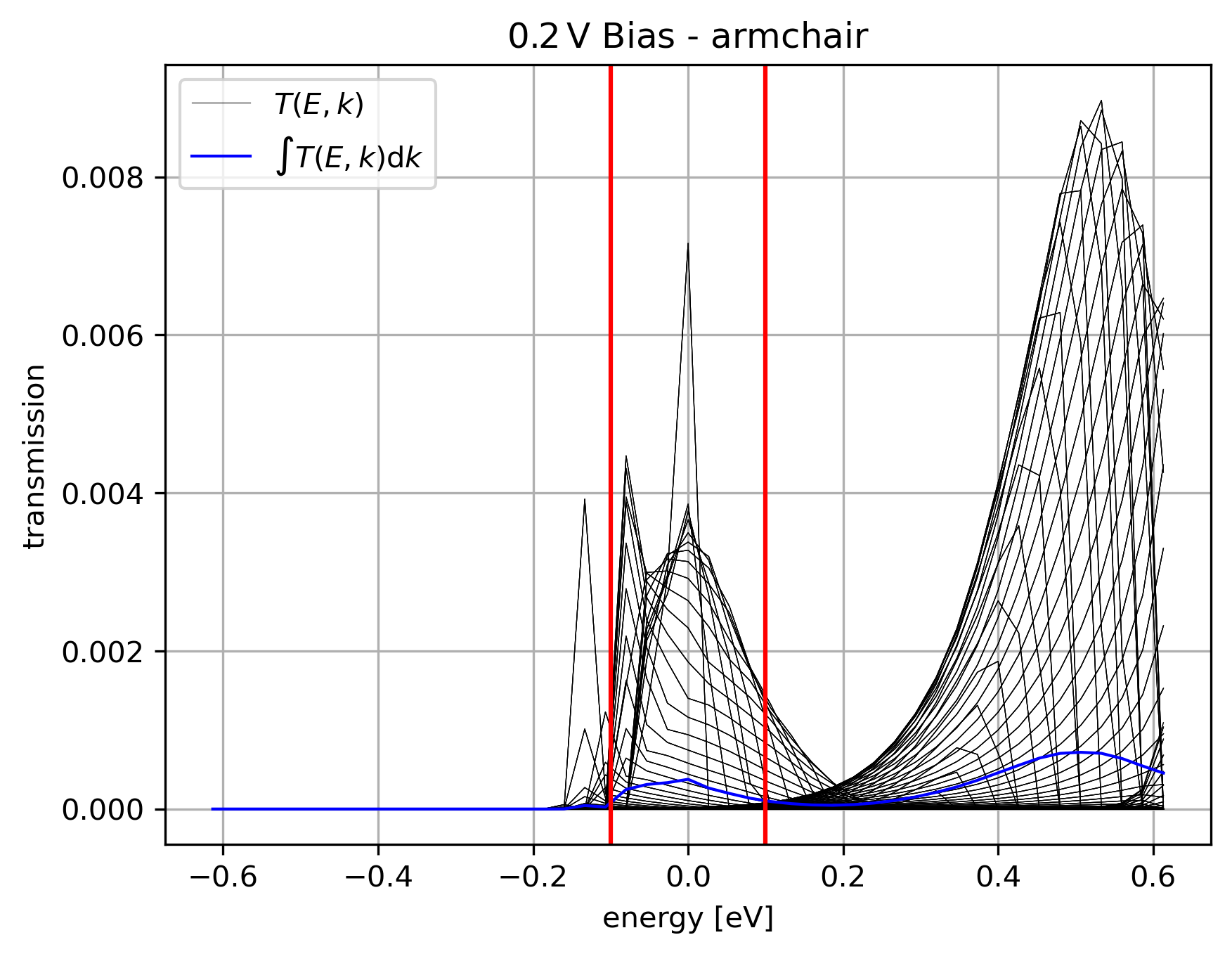}
        \caption*{\SI{0.2}{V} Bias}
    \end{minipage}
    \hfill
    \begin{minipage}{0.32\textwidth}
        \includegraphics[width=\textwidth]{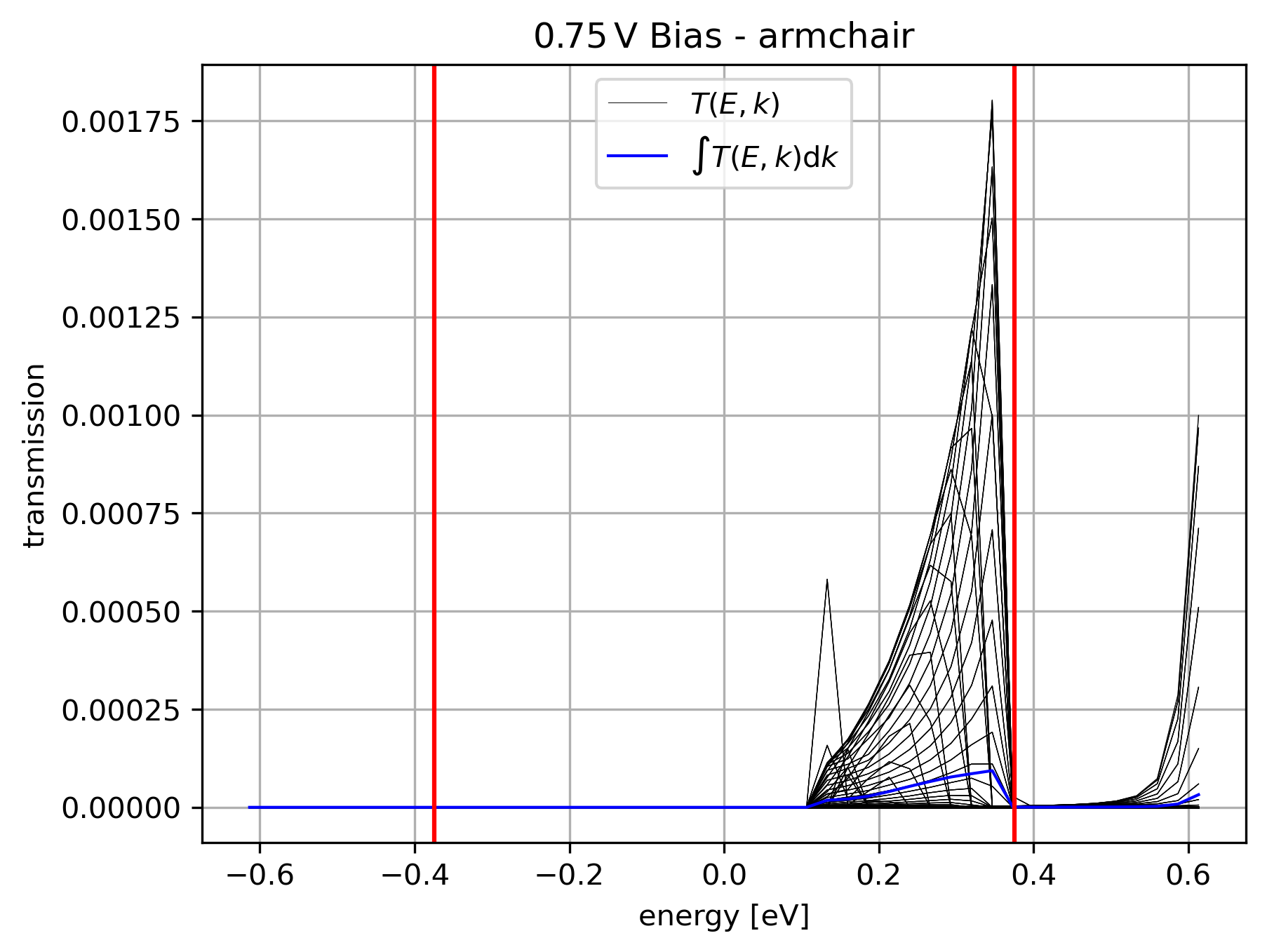}
        \caption*{\SI{0.75}{V} Bias}
    \end{minipage}
    \caption{$k$-point- and energy-resolved transmission spectrum of the armchair device with \SI{21}{\angstrom} barrier for different bias voltages. The vertical red lines show the bias voltage window.
    \label{fig:suppl_fig7}}
\end{figure}

\begin{figure}[H]
    \centering
    \begin{minipage}{0.32\textwidth}
        \includegraphics[width=\textwidth]{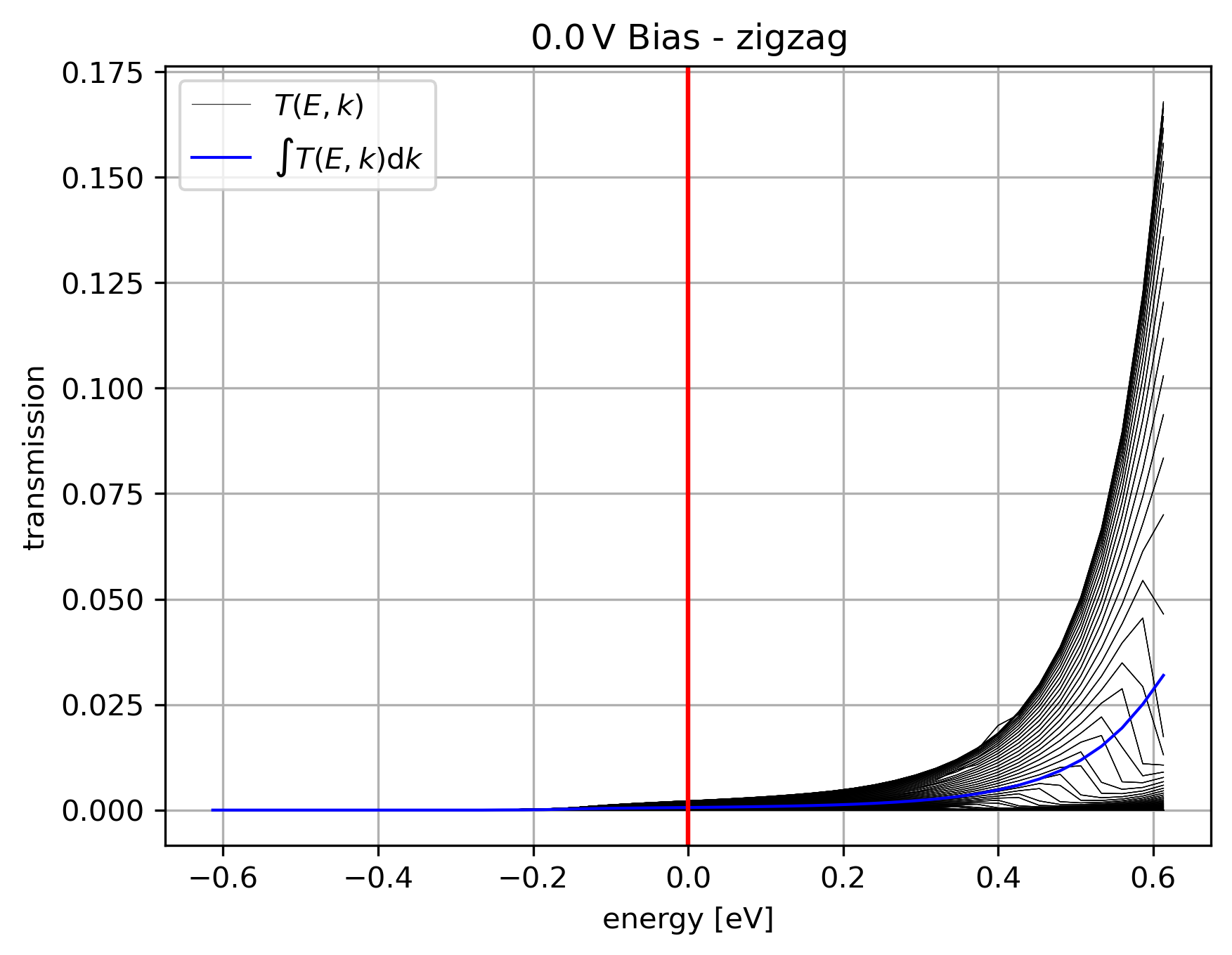}
        \caption*{\SI{0.0}{V} Bias}
    \end{minipage}
    \hfill
    \begin{minipage}{0.32\textwidth}
        \includegraphics[width=\textwidth]{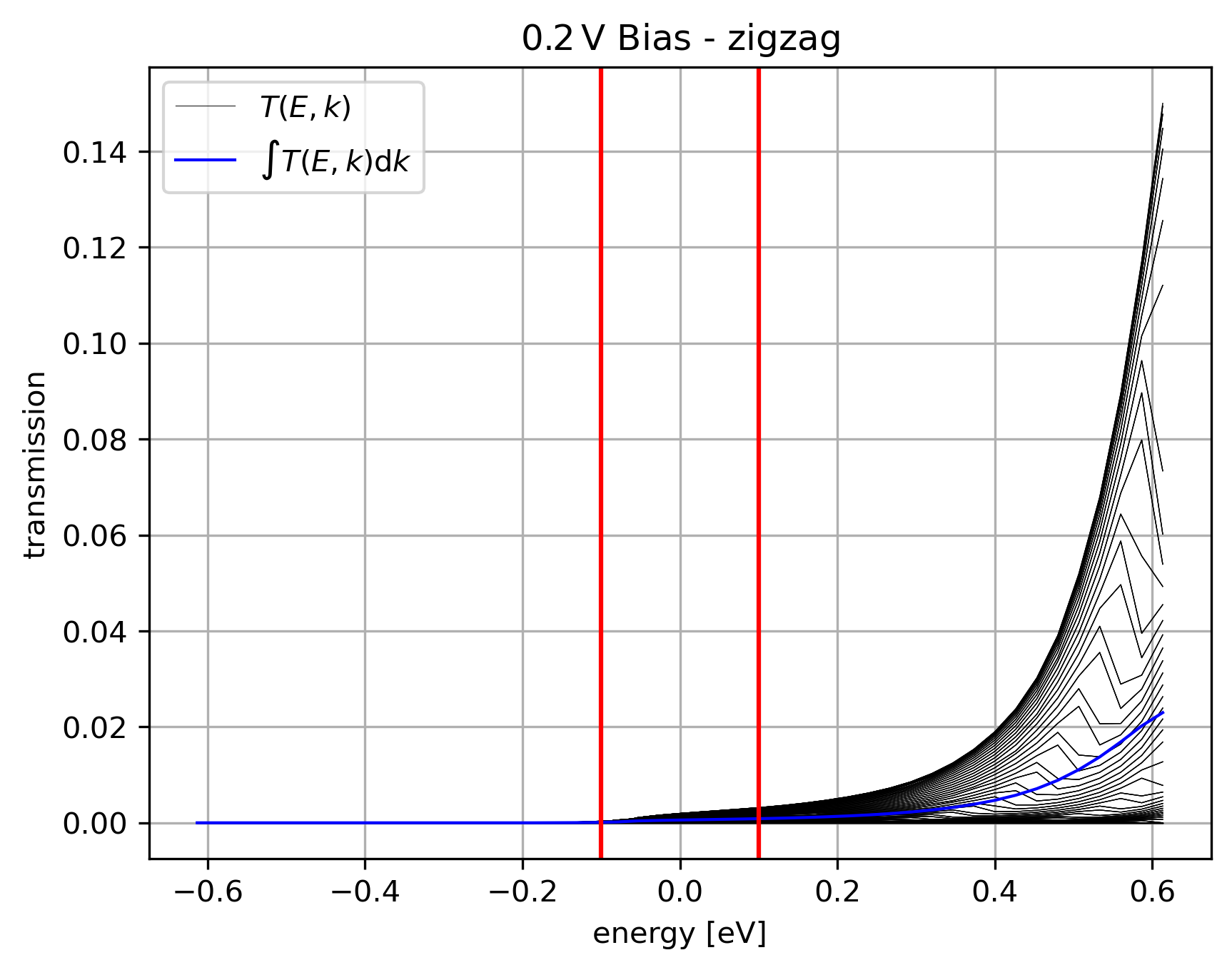}
        \caption*{\SI{0.2}{V} Bias}
    \end{minipage}
    \hfill
    \begin{minipage}{0.32\textwidth}
        \includegraphics[width=\textwidth]{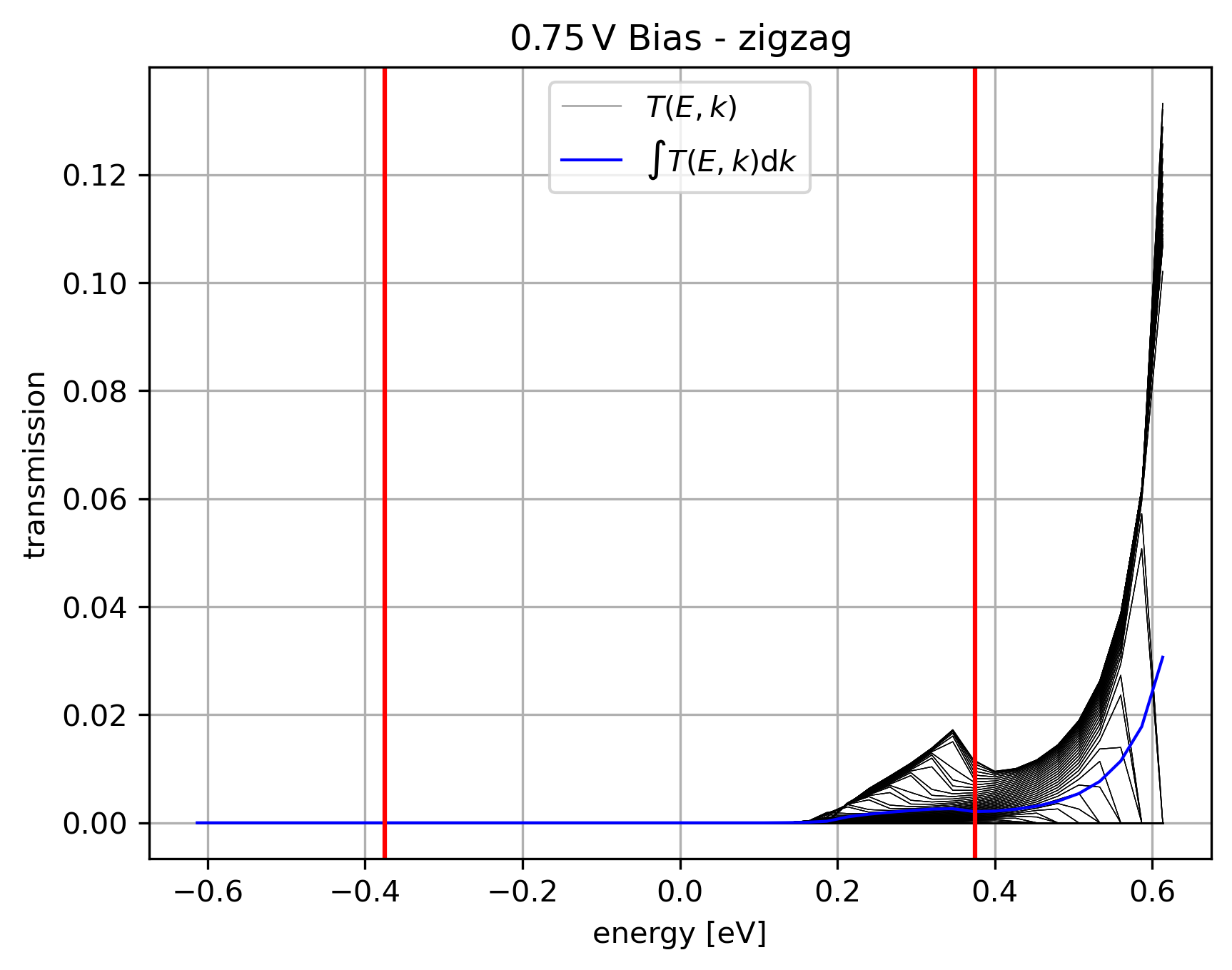}
        \caption*{\SI{0.75}{V} Bias}
    \end{minipage}
    \caption{$k$-point- and energy-resolved transmission spectrum of the zigzag device with \SI{21}{\angstrom} barrier for different bias voltages. 
    The vertical red lines show the bias voltage window.
    \label{fig:suppl_fig8}}
\end{figure}

In Fig.~\ref{fig:suppl_fig7} and Fig.~\ref{fig:suppl_fig8} the $k$- and energy-resolved transmission spectrum is shown for three different
bias voltages. Each black curve represents the transmission for a single $k$-point, and the red lines show the bias window 
($\left[-\frac{V_{bias}}{2}, \frac{V_{bias}}{2}\right]$). In the following, we will call $T(E)$ the transmission and $T(E, k)$ the $k$-resolved 
transmission. Since we are calculating the current within the Landauer-Büttiker approach, the main quantity is the transmission, which can be calculated 
by integrating $T(E,k)$ over all $k$-points. This is shown by the blue curve. Finally, one obtains the $I$-$V$-characteristics by integrating $T(E)$ over a broad energy range. Here, we show $k$- and $E$-resolved transmissions for the \SI{21}{\angstrom}-barrier armchair (Fig.~\ref{fig:suppl_fig7}) and the \SI{21}{\angstrom}-barrier zigzag device (Fig.~\ref{fig:suppl_fig8}). The bias voltages of \SI{0.0}{V}, \SI{0.2}{V}, and \SI{0.75}{V} were chosen, as the maximum of the current in the armchair device occurs at \SI{0.2}{V} and the maximum of the zigzag around \SI{0.7}{V}. For the armchair device, we deduce that the area under the transmission curve is significantly higher at \SI{0.2}{V}  than for \SI{0.75}{V}, underpinning the maxima in the $I$-$V$-curves (fig.~\ref{fig:suppl_fig5}). For the zigzag device, we spot an increase in the area under the transmission curve for the higher bias voltages. Furthermore, one notices that the areas under the transmission curves for the armchair device are much smaller than for the zigzag device, supporting the difference in the current densities absolute values for both devices.

\begin{figure}[H]
    \centering
    \includegraphics[width=0.6\textwidth]{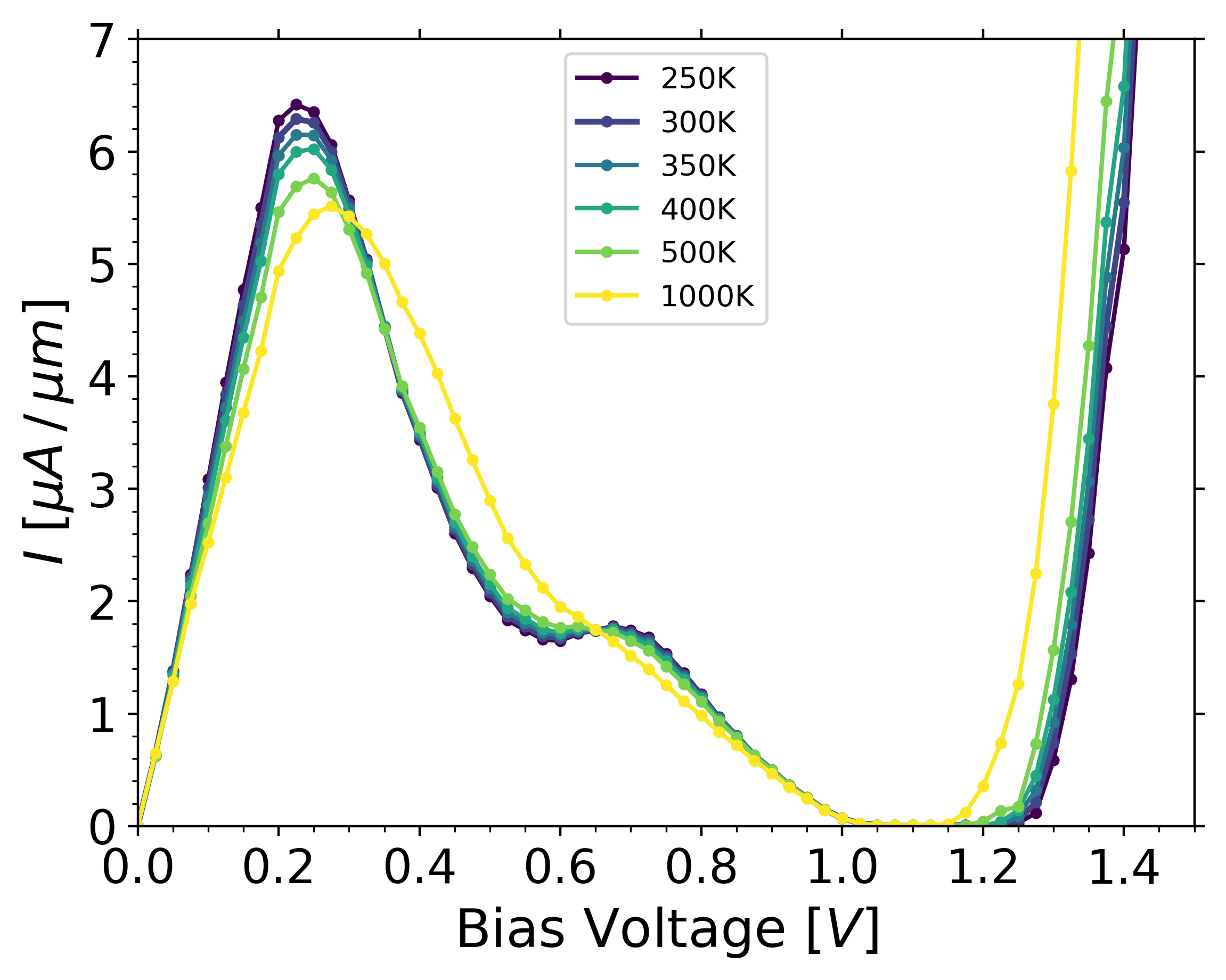}
    \caption{Temperature dependence of the armchair-device with the \SI{21}{\angstrom} barrier. An increase in temperature results in a slight decrease in the peak current as well as a smaller current valley region.}
    \label{fig:temperature}
\end{figure}

\end{document}